\documentclass[twocolumn]{aastex7} 
\usepackage{epsfig}
\usepackage{graphicx} 
\usepackage{float} 
\usepackage{amsmath}
\usepackage{color}
\usepackage{amssymb}
\usepackage{amsfonts} 
\usepackage{units} 
\usepackage{bm}
\usepackage{longtable} 
\usepackage{booktabs}
\usepackage{multirow}
\usepackage[mathscr]{euscript} 
\usepackage{soul}
\usepackage{listings}
\usepackage{algorithm}
\usepackage{algpseudocode}
\usepackage{tikz}
\usetikzlibrary{shapes.geometric, arrows.meta, positioning, calc}

\newcommand\be{\begin{eqnarray}}
\newcommand\ee{\end{eqnarray}}

\begin{document}

\title{A Generalized Algorithmic Framework for Detecting Faraday Rotation Measure Flares in Repeating Fast Radio Bursts}

\correspondingauthor{Yuan-Pei Yang (ypyang@ynu.edu.cn)}

\author[0000-0001-6374-8313]{Yuan-Pei Yang}
\affiliation{South-Western Institute for Astronomy Research, Yunnan Key Laboratory of Survey Science, Yunnan University, Kunming, Yunnan 650504, People's Republic of China}
\email{ypyang@ynu.edu.cn} 

\author[0009-0007-3308-6962]{Boyang Liu}
\affiliation{South-Western Institute for Astronomy Research, Yunnan Key Laboratory of Survey Science, Yunnan University, Kunming, Yunnan 650504, People's Republic of China}
\email{2291069316@qq.com}  

\begin{abstract}

Variations in the Faraday rotation measure (RM) of repeating fast radio bursts (FRBs) provide critical diagnostics of the dynamically evolving magneto-ionic environments surrounding their progenitors. Sudden, transient ``RM flares'' can trace the passage of discrete magneto-ionic structures, such as stellar coronal mass ejections from the companion or other dense plasma clumps, across the line of sight. However, identifying these rare events is difficult because RM evolution manifests a wide range of complex behaviors, from smooth, long-term trends to chaotic stochasticity, further complicated by highly non-uniform temporal sampling. This complexity makes it a non-trivial challenge to distinguish localized ``flares'' from intrinsic environmental volatility. We present a generalized algorithmic framework that establishes a robust methodology for the automated detection and characterization of RM flares. By isolating discrete transient perturbations from quiescent backgrounds, this pipeline enables the uniform census of environmental variability across the FRB population. Applying this framework to 15 repeating FRBs, we find that distinct RM flares are rare, with FRB 20220529A being the only source to exhibit an algorithmic detection under standardized parameters. Most of other active repeaters instead display high-level intrinsic fluctuations or secular evolution. This work provides a rigorous foundation for distinguishing between different modes of local plasma dynamics, offering a crucial diagnostic tool for identifying the diverse progenitor systems and local environments of FRBs.

\end{abstract}


\section{Introduction}

Fast radio bursts (FRBs) are millisecond-duration radio transients of extragalactic origin, characterized by extremely high brightness temperatures. Since the discovery of the inaugural burst \citep{Lorimer07}, the catalog of these events has grown to include over 4500 sources, nearly one hundred of which exhibit repeating behavior \citep{CHIME21,CHIME26}. The identification of FRB 20200428 from the Galactic magnetar SGR 1935+2154 provided pivotal evidence that at least a subset of FRBs can originate from magnetars \citep{CHIME20,Bochenek20,Mereghetti20,Li21,Ridnaia21,Tavani21}. However, the observation of sources in ancient stellar populations, such as globular clusters \citep{Bhardwaj21,Kirsten22}, hints that the progenitor population may be diverse, potentially including systems formed via compact binary mergers \citep{Wang16,Zhang20,Lu22} or other alternative channels \citep[see the review of][]{Zhang23}.

To distinguish between these progenitor models, the propagation effects imprinted on the radio signal serve as critical diagnostic tools. The dispersion measure (DM) and the Faraday rotation measure (RM) are the two most essential parameters for characterizing the plasma environment along the line of sight. While the DM traces the total column density of free electrons, the RM is more sensitive to the magneto-ionic environment surrounding the FRB central engine, being defined as ${\rm RM} \propto \int n_{e} B_{\parallel} dl$, where $n_{e}$ is the electron density and $B_{\parallel}$ is the magnetic field component parallel to the line of sight. Given that the contributions from the intergalactic medium and the Milky Way are often relatively stable or small, a high or rapidly varying RM is generally interpreted as a signature of a dense, highly magneto-ionic environment in the immediate vicinity of the FRB source \citep{Yang23}.

Monitoring the temporal evolution of the RM in repeating FRBs has revealed a high degree of complexity and dynamical change. For instance, the long-term decrease in RM observed in FRB 20121102A was initially attributed to the expansion of a young supernova remnant or a magnetar nebula \citep{Piro18}. In contrast, other repeaters like FRB 20190520B have shown dramatic sign reversals \citep{Anna-Thomas23}, suggesting a more chaotic changing magneto-ionic environment. These variations imply that the local environments of FRBs are not static but are instead shaped by powerful outflows, winds, or orbital motions that continuously reconfigure the RM \citep{Yang23}. 

A significant discovery was very recently reported by \citet{Li26}, in which work they observed a dramatic and rapid transformation in the RM evolution of the repeating source FRB 20220529A. While the source had maintained a relatively stable RM with a median of approximately 17 rad m$^{-2}$ for 17 months, it exhibited a sudden jump to nearly $2000~{\rm rad~m^{-2}}$ in December 2023, followed by a recovery to typical levels within just two weeks. This phenomenon, which was termed as the ``RM flare'', represents a sudden, localized excursion in the RM that provides a direct diagnostic of transient, high-density magneto-ionic structures passing through the line of sight. 
Such events are of significant scientific interest because they likely trace discrete plasma clumps, potentially associated with coronal mass ejections from a companion star in a binary system or extreme localized dense plasma within a surrounding nebula. Consequently, the detection and systematic analysis of RM flares offer a unique method to probe small-scale magnetic field fluctuations and provide indirect evidence for the presence of companion objects.

Despite the high diagnostic potential of RM flares, their discovery has remained largely opportunistic, often relying on manual inspection of specific active episodes. Identifying discrete flares within RM time-series is significantly more complex than detecting signals against stationary noise because FRB RM evolution manifests a wide range of behaviors. These patterns range from smooth, long-term secular trends to chaotic stochasticity. When these diverse behaviors are coupled with highly non-uniform temporal sampling, the isolation of localized physical perturbations becomes highly challenging. Distinguishing a genuine flare caused by a discrete structure from intrinsic environmental fluctuations or long-term baseline evolution requires a sophisticated approach that can adapt to various morphological patterns. As high-cadence monitoring programs from major facilities like FAST, CHIME, and Parkes continue to accumulate an increasing volume of polarization data, the lack of a standardized detection framework has become a significant obstacle. There is a pressing need for a generalized, robust algorithm that can systematically scan RM time-series data to identify flare-like excursions and distinguish them from underlying secular evolution or stochastic noise. Such a tool would enable a more uniform census of environmental volatility across the FRB population. 

In this paper, we develop a generalized algorithmic framework for the automated detection and characterization of RM flares in repeating FRB sources. Our methodology features an adaptive windowing technique designed to establish a stable quiescent baseline from non-uniformly sampled time-series data and utilizes a robust significance-scoring logic to identify transient events. 
We apply this pipeline to a representative sample of 15 repeating FRBs that have exhibited significant RM variability, allowing for a systematic search for flare-like perturbations across the population. 
The paper is organized as follows. In Section \ref{methods}, we detail the mathematical architecture and implementation of the detection pipeline. In Section \ref{results}, we present the results of the systematic census across the FRB sample. Finally, the results are discussed and summarized in Section \ref{conclusions}.

\section{Methods}\label{methods}

\begin{figure*}[ht!]
    \centering
    \tikzset{
        base/.style = {rectangle, rounded corners, draw=black, text width=7.2cm, align=center, font=\sffamily\small, thick, inner sep=5pt},
        input/.style = {base, fill=blue!5},
        process/.style = {base, fill=orange!8},
        loop_proc/.style = {base, fill=green!8},
        decision/.style = {diamond, aspect=2.0, draw=black, fill=yellow!15, align=center, font=\sffamily\small, thick, inner sep=1pt, text width=2.8cm},
        output/.style = {base, fill=red!10},
        arrow/.style = {thick, ->, >=stealth}
    }
    \resizebox{0.85\textwidth}{!}{
    \begin{tikzpicture}
        \node (in) at (0, 0) [input] {\textbf{Input:} Raw Non-uniform RM Time Series \\ $\{t, {\rm RM}(t), \sigma_{\rm err}\}$};
        \node (stage1) at (0, -2.6) [process] {\textbf{Stage 1: Adaptive Windowing \& Initialization} \\ Extract unique epochs to determine window $W$ \\ Calculate global median and $\text{MAD}_{\rm glob}$ \\ Mask extreme outliers $> N_{\rm glob}\text{MAD}_{\rm glob} + \Delta_{\rm RM}$};
        \node (stage2) at (0, -5.8) [loop_proc] {\textbf{Stage 2: Iterative Baseline Estimation} \\ Calc rolling median on quiescent subset \\ Interpolate $B_{\rm temp}(t), \sigma_{\rm temp}(t)$ to all epochs $t$ \\ Filter points $< N_{\rm loc}\sigma_{\rm temp}(t) + \delta_{\rm RM}$ and exclude extremes}; 
        \node (dec1) at (0, -8.6) [decision] {Quiescent Mask \\ Converged?};
        \node (stage3) at (8.5, -2.6) [process] {\textbf{Stage 3: Multi-Component Noise \& Scoring} \\ Total noise $\sigma_{\rm tot}(t) = \sqrt{\sigma_{\rm glob}^2 + \sigma_{\rm loc}(t)^2 + \sigma_{\rm intra}(t)^2 + \sigma_{\rm err}(t)^2}$ \\ Significance Score $S(t) = \mathcal{R}(t) / \max(\sigma_{\rm safe}, \sigma_{\rm tot}(t))$};
        \node (dec2) at (8.5, -5.8) [decision] {Peak Score \\ $S_{\rm peak} \ge T_{\rm tri}$ ?};
        \node (stage4) at (8.5, -8.6) [output] {\textbf{Stage 4: Flare Phase Bounding} \\ Interpolate exact timestamps where $S(t) = \eta S_{\rm peak}$ \\ to define continuous boundaries $[t_{\rm start}, t_{\rm end}]$};
        \node (out) at (8.5, -11.0) [input] {\textbf{Output:} Adaptive Baseline $B(t)$, Scores $S(t)$, \\ and Validated Flare Intervals};
        \draw [arrow] (in) -- (stage1);
        \draw [arrow] (stage1) -- (stage2);
        \draw [arrow] (stage2) -- (dec1);
        \draw [arrow] (stage3) -- (dec2);
        \draw [arrow] (stage4) -- (out);
        \draw [arrow] (dec1.west) -- (-4.5, -8.6) |- node[anchor=east, pos=0.25, font=\sffamily\bfseries] {No} (stage2.west);
        \draw [arrow] (dec1.east) -- node[anchor=south, font=\sffamily\bfseries] {Yes} (4.25, -8.6) |- (stage3.west);
        \draw [arrow] (dec2.east) -- (13.0, -5.8) |- node[anchor=west, pos=0.25, font=\sffamily\bfseries] {No} (out.east);
        \draw [arrow] (dec2.south) -- node[anchor=west, font=\sffamily\bfseries] {Yes} (stage4.north);
    \end{tikzpicture}
    }
    \caption{
    Architecture of the Automated RM Flare Detection Pipeline. The framework is driven by four core components: adapting to observational cadence directly from raw timestamps (Stage 1), robustly isolating the quiescent background through an iterative loop featuring in-loop interpolation to full resolution (Stage 2), constructing a multi-component pseudo-SNR metric (Stage 3), and dynamically defining the physical boundaries of discrete flares (Stage 4). The code is available at Science Data Bank: \dataset[doi: 10.57760/sciencedb.36845]{\doi{10.57760/sciencedb.36845}}, and the corresponding algorithm pseudocode is shown in Appendix \ref{pseudocode}. 
    }\label{fig:flowchart}
\end{figure*}

The identification of RM flares requires a computational framework capable of distinguishing transient, high-amplitude excursions from both the slowly evolving magneto-ionic background and the stochastic noise inherent in non-uniformly sampled observations. To achieve this, in this section we develop a data-driven algorithm that incorporates adaptive temporal windowing and iterative baseline estimation. Specifically, the procedure is designed to mitigate signal self-subtraction during baseline fitting. The primary objective of this framework is to provide a standardized approach to systematically isolate discrete flares from the overall RM evolution observed in repeating FRBs. 
The flowchart and the pseudocode of our automated RM flare detection pipeline are shown in Figure \ref{fig:flowchart} and in Appendix \ref{pseudocode}, respectively.

\subsection{Adaptive Baseline Estimation: Quiescent Background Modeling}\label{baseline}

A primary challenge in analyzing repeating FRB data is the irregular observational cadence, where the interval between observing sessions can range from hours to several months. A fixed smoothing timescale is fundamentally unsuitable for such datasets; a narrow window would lead to an unstable, noise-driven baseline in sparse records, while a wide window would over-smooth the intricate secular details in high-cadence observations. 
To maintain statistical consistency across disparate datasets, our algorithm implements an adaptive temporal window size, $W$, which scales according to the typical observational cadence of each source. The adaptive window $W$ is implemented as a time-duration-based rolling window rather than a fixed number of samples, ensuring robustness against highly non-uniform sampling. 

To ensure the stability of the following calculations, the algorithm first performs error preprocessing by imputing any missing or negative reported instrumental errors using the median value of the strictly positive error distribution in the dataset. In extreme cases where no valid positive errors exist within the entire dataset, a static fallback value of $1.0~{\rm rad~m^{-2}}$ is adopted. To gauge the underlying observational cadence, we extract the set of unique observing epochs (Modified Julian Dates, MJDs) and calculate the median time interval between consecutive unique observations, $\Delta t_{\rm med}$. If only a single unique observing epoch exists in the dataset, a default interval of $\Delta t_{\rm med} = 10$ days is assigned. Operating directly on the raw, full-resolution timeline rather than artificially down-sampling or aggregating the data, the algorithm defines an adaptive temporal window size. This window is calculated by scaling the median gap $\Delta t_{\rm med}$ by a multiplier $k_w$ (default $k_w = 25$) and applying a boundary constraint through a clip operator: 
\begin{equation}
    W = \text{clip}(k_w\Delta t_{\rm med}, W_{\rm min}, W_{\rm max}),
\end{equation}
where $W_{\rm min}$ and $W_{\rm max}$ represent the lower and upper bounds of the adaptive window, respectively. In our default configuration, we set $W_{\min} = 30$ days and $W_{\max} = 150$ days. 
The function ${\rm clip}(x, a, b)$ is defined as a clipping operator that restricts the input $x$ to the range $[a, b]$, such that: 
\begin{equation}
{\rm clip}(x, a, b) = 
\begin{cases} 
a & \text{if } x < a, \\
x & \text{if } a \le x \le b, \\
b & \text{if } x > b.
\end{cases}
\end{equation}

The scientific rationale for this bounded adaptation is to ensure the baseline remains robust while staying physically responsive. The lower bound, $W_{\rm min}$, ensures that even for high-cadence data, the window remains large enough to prevent the baseline from ``over-fitting'' to short-term noise fluctuations. Conversely, the upper bound, $W_{\rm max}$, ensures that in sparse datasets, the baseline does not become so rigid that it fails to track genuine long-term RM evolution, such as that produced by binary orbital motion or the expansion of a local supernova remnant. By anchoring the baseline estimation to a timescale that typically encompasses $k_w=25$ individual observations, the algorithm provides a continuous and reliable reference of the stable RM evolution. 

Furthermore, a critical challenge is the prevention of signal self-subtraction. If data points belonging to an RM flare are included in the baseline calculation, the baseline itself will shift toward the flare, thereby artificially reducing the flare's significance and distorting its profile. To isolate the quiescent background, we implement a two-stage iterative rejection process that relies on data-driven parameters rather than fixed, hard-coded thresholds. First, we automatically extract adaptive bias constraints based on the intrinsic distribution of the dataset. Specifically, the global bias offset ($\Delta_{\rm RM}$) is derived from the interquartile range (IQR) of the RM values to accommodate the overall dispersion, defined as $\Delta_{\rm RM} = \max(1.5\times {\rm IQR}, 20.0~{\rm rad~m^{-2}})$. 
Similarly, the local bias offset ($\delta_{\rm RM}$) is derived from the step-wise differences between consecutive epoch observations, defined as $\delta_{\rm RM} = \max(2.0\times \text{Median}(|\Delta {\rm RM}_{i,i-1}|), 5.0~{\rm rad~m}^{-2})$.
Finally, a dynamic noise floor ($\sigma_{\rm floor}$) is established to prevent over-fitting in regions with underestimated instrumental errors, defined as $\sigma_{\rm floor} = \max({\rm Median}(\sigma_{\rm err}), 2.0~{\rm rad~m^{-2}})$.

In the first stage (global screening), evaluated on the epoch series, we mask extreme outliers that strictly exceed the global median by more than $N_{\rm glob}\text{MAD}_{\rm glob}+\Delta_{\rm RM}$, where $\text{MAD}_{\rm glob}$ is the global Median Absolute Deviation. We adopt $N_{\rm glob} = 10$ (default). This ensures that high-amplitude flares are excluded from the initial trend estimation.

Second, we apply a local robust filter via an iterative smoothing procedure within the adaptive window $W$. In each iteration, a rolling median is applied to the surviving quiescent epochs to construct a temporary baseline $B_{\rm temp}(t)$. The localized noise, $\sigma_{\rm temp}(t)$, is then empirically estimated by calculating the rolling median of the absolute residuals around this temporary baseline, constrained by the minimum floor $\sigma_{\rm floor}$. Epochs are then re-evaluated: a new quiescent mask is formed by selecting points whose absolute deviation from the localized baseline is less than $N_{\rm loc}\sigma_{\rm temp}(t) + \delta_{\rm RM}$ (default $N_{\rm loc} = 5$), while excluding the previously identified global extreme outliers. This intersection ensures mathematical convergence without entering a logic loop. This process is iterated (up to 10 times) until the quiescent mask mathematically converges.  

Within each iteration of the above process, both the temporary baseline $B_{\rm temp}(t)$ and the localized noise $\sigma_{\rm temp}(t)$ are linearly interpolated to map back onto the original high-resolution temporal timestamps $t$. Upon convergence, this yields a continuous quiescent baseline $B(t)$ and a localized noise profile $\sigma_{\rm loc}(t)$ across observation gaps. This data-driven approach ensures a ``clean'' reference that is resilient to localized flares, non-uniform sampling, and widely varying source-intrinsic noise characteristics. 

\subsection{Significance Scoring and Flare Detection}

To robustly quantify the prominence of an observed deviation against the dynamic background, a fundamental challenge must first be addressed: due to the non-stationary, non-Gaussian nature of RM evolution and the highly non-uniform observational cadence, formulating a rigorous likelihood framework or a formal statistical hypothesis test is computationally and physically intractable. To overcome these limitations, we define an empirical metric termed the ``Significance Score'', $S(t)$. Rather than serving as a strict probabilistic statistic associated with a known false-alarm probability, $S(t)$ functions as an empirical, non-parametric pseudo-Signal-to-Noise Ratio (SNR). Specifically, it evaluates significance against a noise model that incorporates not only the instrumental errors, but also the intrinsic global volatility, local tracking variance, and intra-day scatter.

We first establish the global residual noise, $\sigma_{\rm glob}$, derived from the residuals between the raw RM data and the adaptive baseline $B(t)$. For each data point, the residual is defined as $\mathcal{R}(t) = |{\rm RM}(t) - B(t)|$. To ensure statistical robustness against outliers and flare signatures, we define $\sigma_{\rm glob}$ as the Median Absolute Deviation (${\rm MAD}$) of these residuals:
\begin{equation}
\sigma_{\rm glob} = \text{Median} \left( \left| \mathcal{R}(t) - \text{Median}(\mathcal{R}(t)) \right| \right).
\end{equation}
This global component represents the intrinsic volatility floor of the source over the entire observing span. 

Crucially, to prevent localized flares from polluting the local noise estimation (signal self-subtraction), the local noise component, $\sigma_{\rm loc}(t)$, utilizes the interpolated profile derived from the robust iterative filtering in Section \ref{baseline}. Rather than using the standard deviation of raw data, this component actively tracks the dynamically varying environment strictly calculated from the quiescent mask.

Furthermore, repeating FRB observations frequently capture multiple bursts within a single observing session (typically within one day), which can exhibit significant instantaneous vertical scatter. To prevent these short-timescale, highly scattered micro-fluctuations from being misidentified as secular RM flares, we introduce an intra-day scatter penalty, $\sigma_{\rm intra}(t)$. This is computed as the standard deviation of all RM values occurring within the same integer MJD day. For days with only a single burst detection, $\sigma_{\rm intra}$ defaults to zero.

The total effective noise $\sigma_{\rm tot}(t)$ is defined as the quadrature sum of these four independent components:
\begin{equation}
\sigma_{\rm tot} = \sqrt{\sigma_{\rm glob}^2 + \sigma_{\rm loc}^2 + \sigma_{\rm intra}^2 + \sigma_{\rm err}^2}.
\end{equation}
By incorporating $\sigma_{\rm loc}$ and $\sigma_{\rm intra}$ into $\sigma_{\rm tot}$, the algorithm actively suppresses false positives in regions of high intrinsic fluctuations, turbulent observing days, or instrumental instability. 

To prevent spurious significance inflation in temporally sparse regions with anomalously low formal errors, we enforce a data-driven safety floor on the denominator. This boundary, $\sigma_{\rm safe} = \max(\sigma_{\rm floor}, \delta_{\rm RM})$, is derived from the baseline limits established in Section \ref{baseline}. The significance score is thus defined as:
\begin{equation}
S(t) = \frac{\mathcal{R}(t)}{\max(\sigma_{\rm safe}, \sigma_{\rm tot}(t))}.
\end{equation}
The detection engine follows a grouped-trigger and physical-interpolation logic. First, the algorithm identifies candidate temporal segments where $S(t) > 0.5$, which ensures that even subtle deviations are initially considered. A flare candidate is formally initiated only if a contiguous segment contains at least one point satisfying $S(t) > T_{\rm tri}$, where $T_{\rm tri} = 10$ (default). Rather than being derived from an analytical false-alarm probability integral, this trigger threshold is a conservative, empirically calibrated boundary, because the standard formal errors underestimate the true volatility. The computationally validated choice of $T_{\rm tri} = 10$ effectively suppresses false-positive triggers arising from secular trends or unmodeled red noise. While a lower threshold of $T_{\rm ref}=5$ is provided as a visual reference for significant deviations, the adoption of $T_{\rm tri}=10$ ensures high algorithmic confidence in isolating discrete, high-amplitude events from background stochasticity.

Once a flare is triggered, the precise temporal boundaries of the ``Flare Phase'' are determined using a physical width-scaling criterion. Instead of simply selecting discrete data points, the algorithm calculates the exact time at which the score curve crosses a dynamic boundary defined by the peak intensity. To achieve sub-cadence temporal precision, the precise MJDs where the score curve crosses the dynamic boundary $\eta S_{\rm peak}$ are calculated via linear interpolation. The continuous ``Flare Phase'' is then defined as the temporal interval $[t_{\rm start}, t_{\rm end}]$ satisfying:
\begin{equation}
S(t) \ge \eta S_{\rm peak},
\end{equation}
where $S_{\rm peak}$ is the maximum significance score within that segment and $\eta$ is the width ratio. In our latest implementation, we adopt $\eta = 0.1$, which corresponds to the Full Width at Tenth Maximum (FWTM) criterion. This approach provides a standardized, physically motivated definition of flare duration that is robust against sparse or non-uniform sampling.

\section{Results}\label{results}

\subsection{Quantitative Validation of the Detection Framework}

Prior to applying the detection framework to the observational sample, we quantitatively validated its performance through a series of controlled experiments using simulated RM time series. These tests were designed to systematically evaluate the algorithm's detection sensitivity and robustness across a diverse parameter space, including varying flare amplitudes, durations, and temporal sampling cadences. 

As detailed in Appendix \ref{sec:appendix_mock}, the simulations delineate the operational boundaries of the algorithm under different observational conditions. Specifically, for sufficiently prominent events, the two-stage iterative baseline estimation prevents signal self-subtraction, preserving the empirical profile of the injected flares. Meanwhile, the multi-component noise model successfully suppresses sub-threshold fluctuations. The tests also demonstrate that while the adaptive windowing mechanism extracts stable backgrounds across varying data densities, the algorithm remains inherently conservative: it intentionally withholds triggers for flares that are too weak, excessively brief, or insufficiently sampled (e.g., the sparse dataset scenarios).

\subsection{Parameter Sensitivity Analysis}
\label{sec:sensitivity}

To verify that the automated flare detections are not overly reliant on finely-tuned hyperparameters, we evaluated the sensitivity of the algorithm to its key configuration parameters, including the window scaling multipliers ($k_w$, $W_{\rm min}$, $W_{\rm max}$), the screening thresholds ($N_{\rm glob}$, $N_{\rm loc}$), and the width ratio ($\eta$). Utilizing the simulated datasets, we systematically varied each parameter across a wide dynamic range to determine whether alternative choices would yield fundamentally different flare detection results.

The analysis indicates that the detection pipeline operates stably within a broad parameter space, even when challenged with extreme, broad flare morphologies. Specifically, the peak significance scores consistently exceed the rigorous trigger threshold ($T_{\rm tri} = 10$) across a wide range of adaptive window boundaries ($W_{\rm min}$ and $W_{\rm max}$) and local filters ($N_{\rm loc}$). Expected performance degradation, such as signal self-subtraction or inflated global noise floor estimates, only occurs when parameters are pushed to physically unconstrained extremes (e.g., overly loose global masking thresholds $N_{\rm glob} \ge 12$).  

Furthermore, the variation of the width ratio parameter ($\eta$) shows that the algorithm consistently captures the physical morphology of the transient event. The extracted flare duration scales predictably with $\eta$, accurately matching the injected Full Width at Half Maximum (FWHM) at $\eta=0.5$ and expanding appropriately to cover the full physical interaction phase (FWTM) under our default $\eta=0.1$. 

Overall, the standardized parameters employed in this framework reside within robust operational regimes. Moderate variations in these values do not alter the underlying flare detection outcomes. A detailed discussion of the parameter space and the corresponding sensitivity curves is provided in Appendix \ref{sec:appendix_sensitivity}. 

\subsection{Application to the Repeating FRB Sample}

\begin{figure*}
    \centering
    \newcommand{\figwidth}{0.46\linewidth}   
    \newcommand{\figsep}{\hspace{0.02\linewidth}} 

    \includegraphics[width = \figwidth, trim = 0 0 0 0, clip]{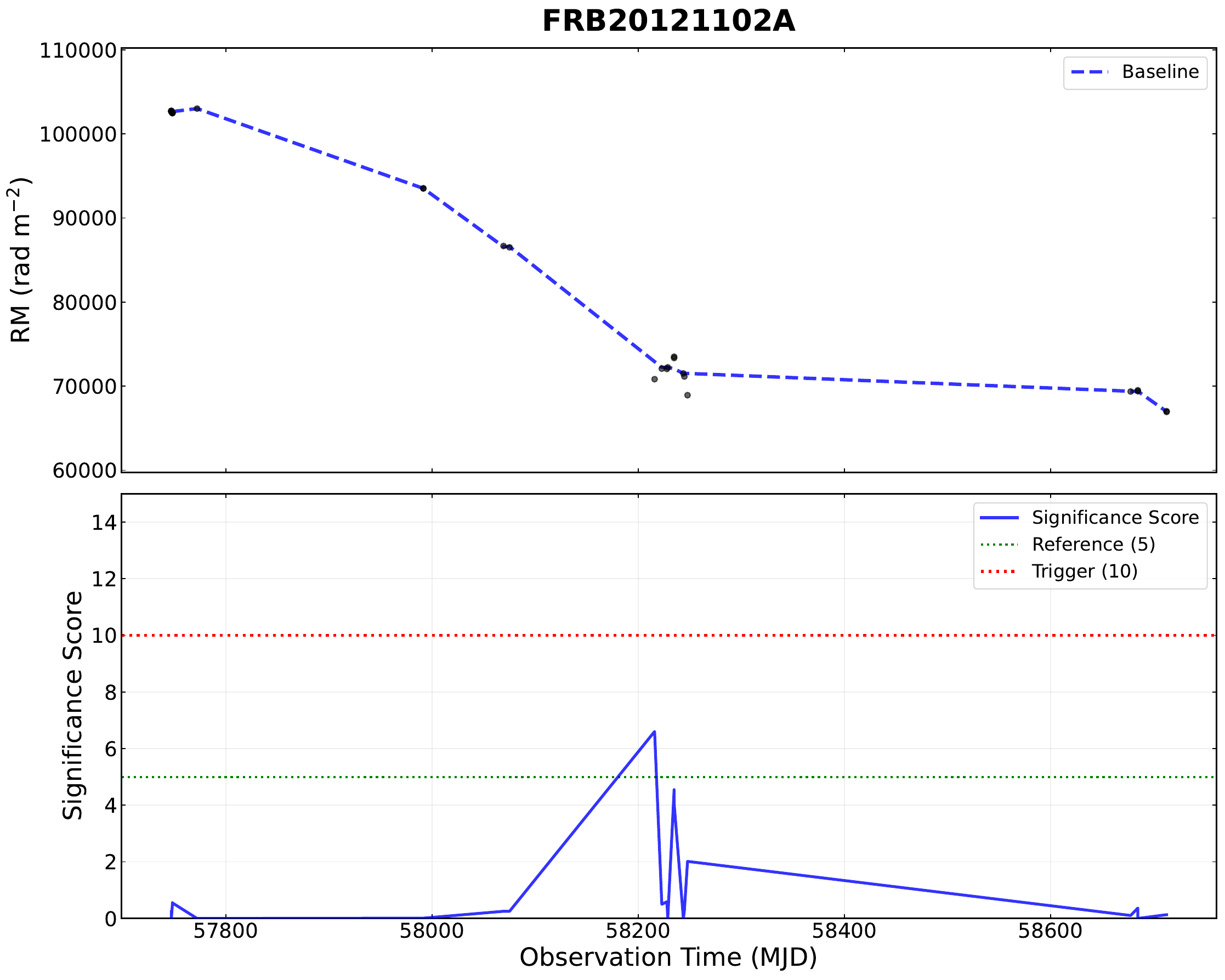} \figsep
    \includegraphics[width = \figwidth, trim = 0 0 0 0, clip]{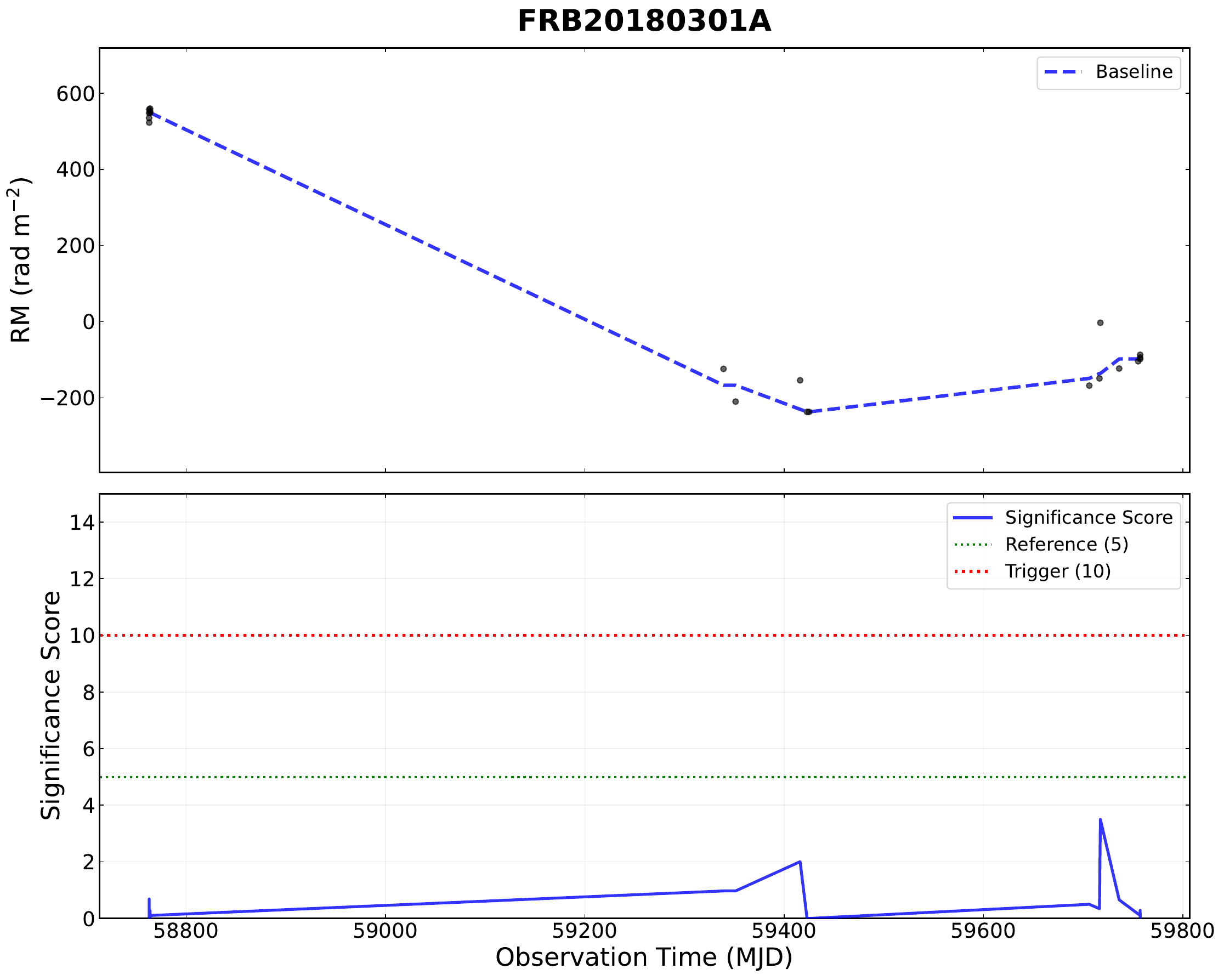} \\ \vspace{0.3cm}
    
    \includegraphics[width = \figwidth, trim = 0 0 0 0, clip]{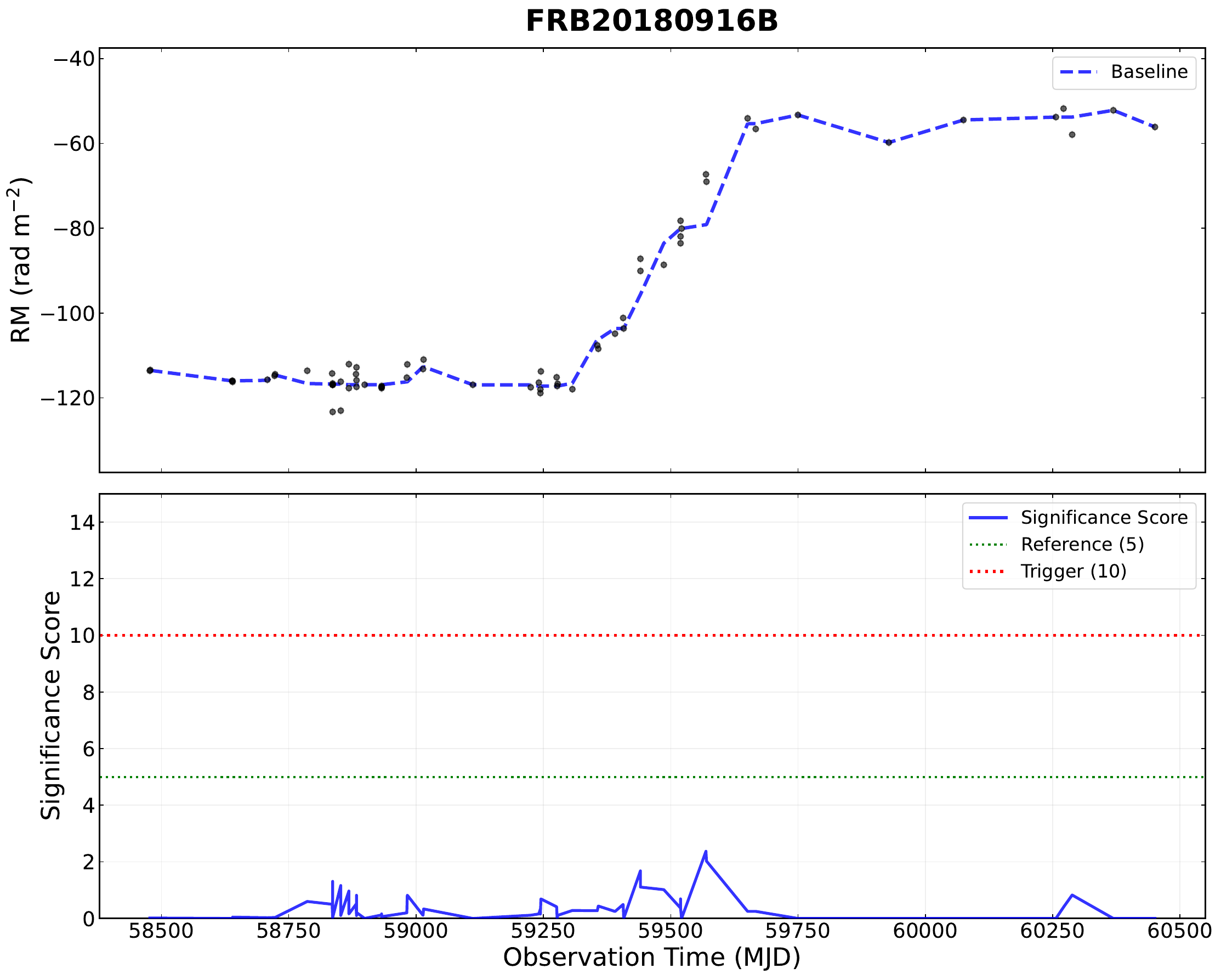} \figsep
    \includegraphics[width = \figwidth, trim = 0 0 0 0, clip]{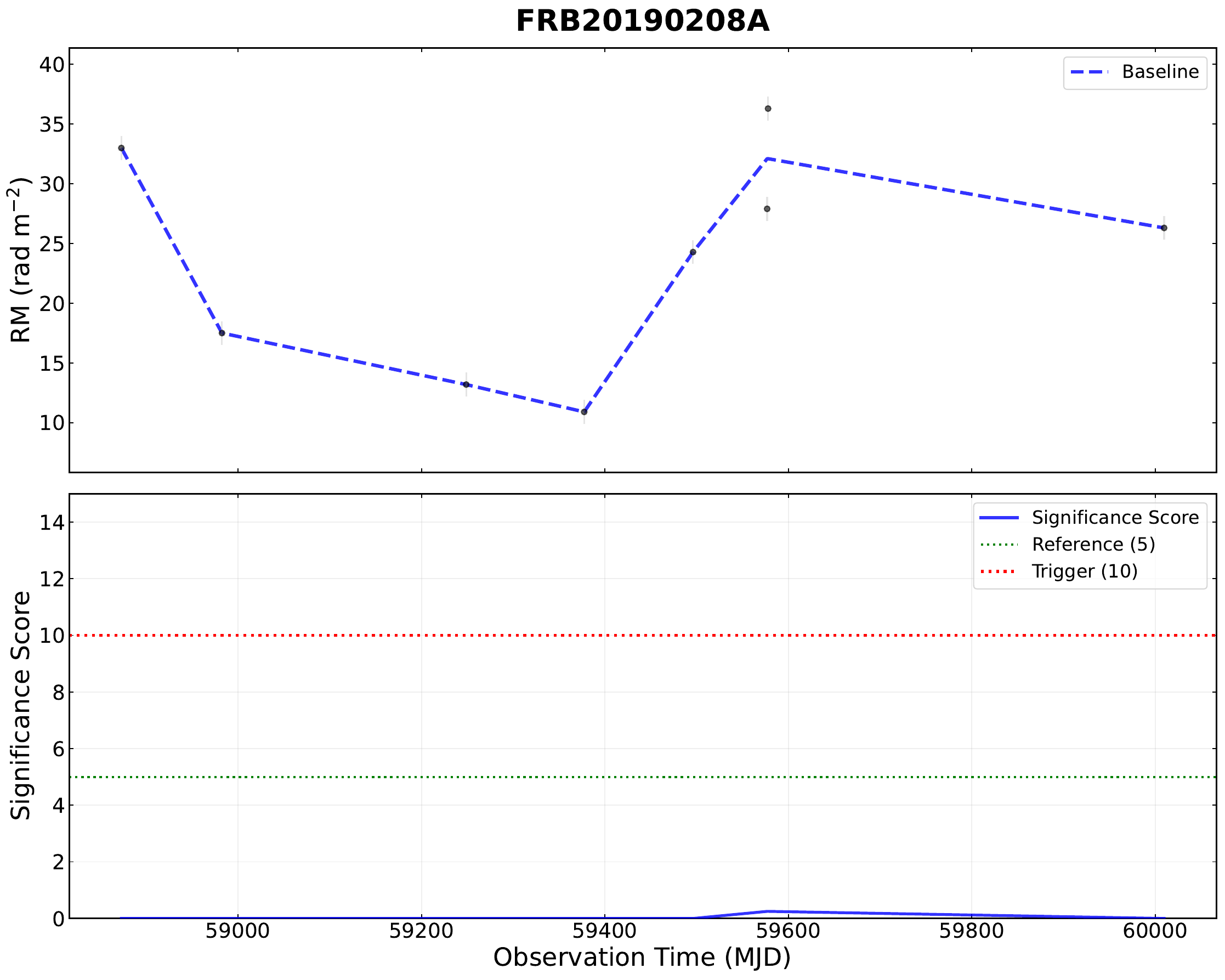} \\ \vspace{0.3cm}
    
    \includegraphics[width = \figwidth, trim = 0 0 0 0, clip]{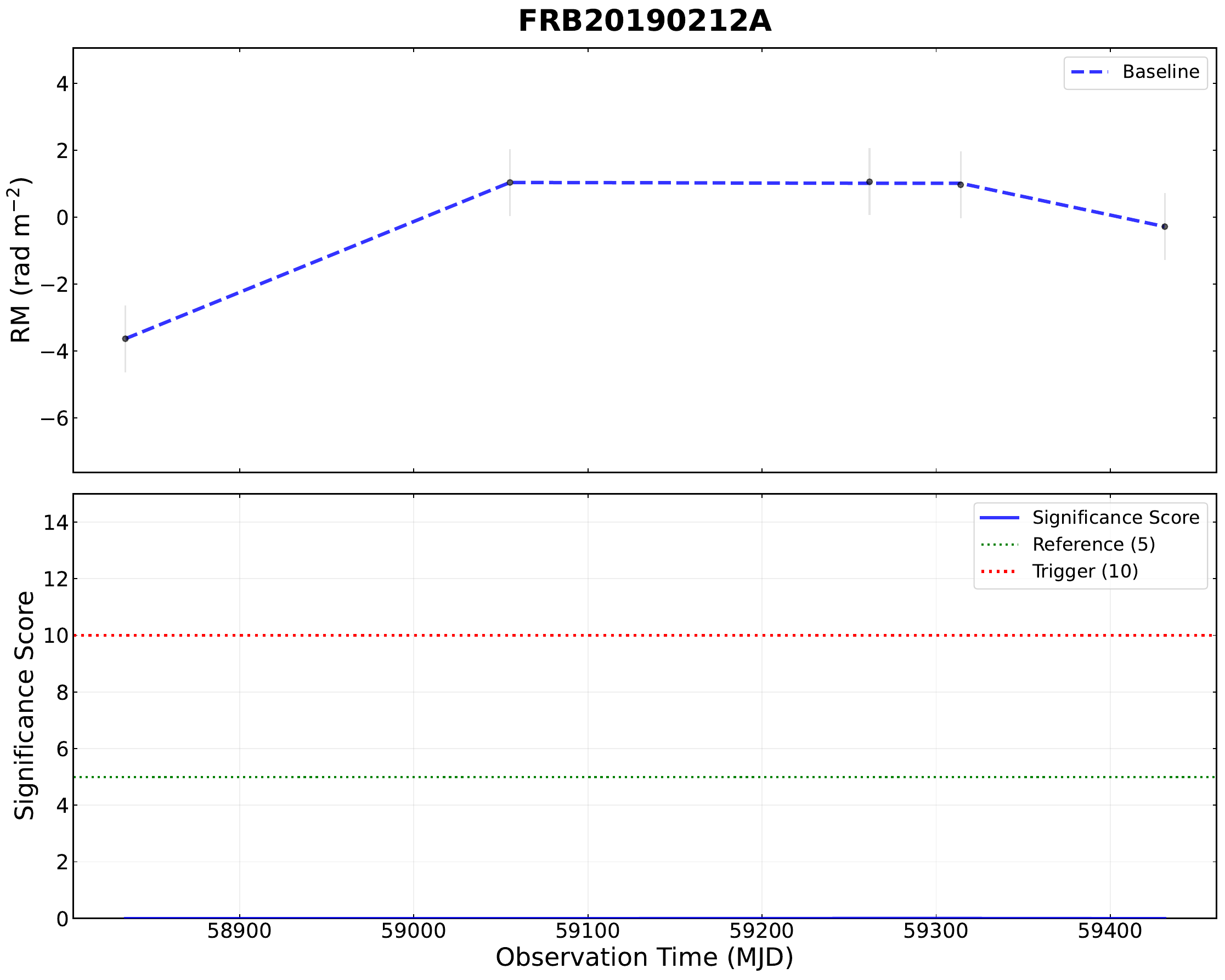} \figsep
    \includegraphics[width = \figwidth, trim = 0 0 0 0, clip]{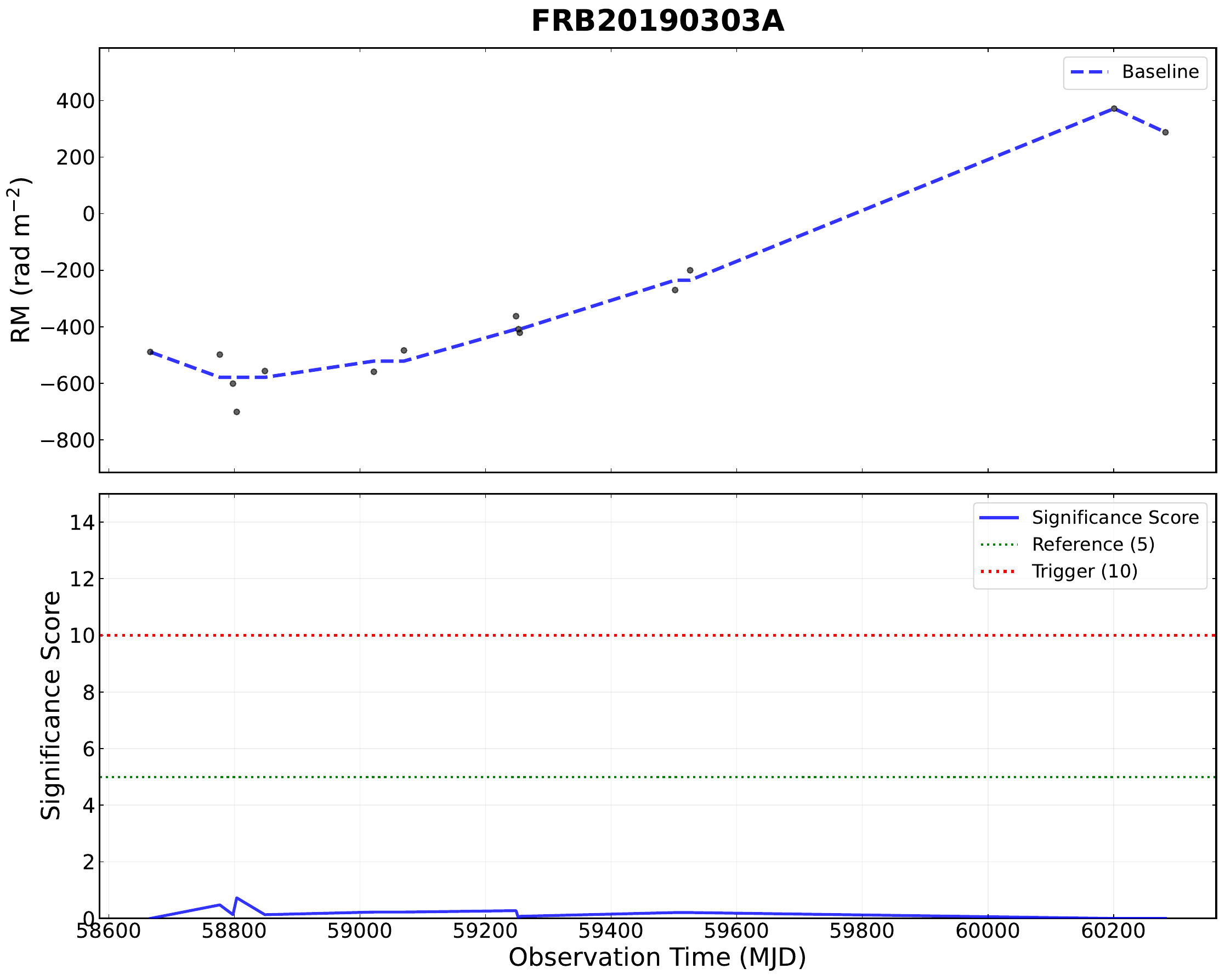}
    
    \caption{Automated flare detection results for 15 repeating FRBs. Top sub-panels: Observed RM (black circles) and the adaptive baseline $B(t)$ (blue dashed line). Bottom sub-panels: Significance Score (blue solid line) relative to the reference threshold (green dotted line, $T_{\rm ref}=5$) and the rigorous trigger threshold (red dotted line, $T_{\rm tri}=10$). Red shaded regions mark the ``Flare Phase'', which is initiated when $S(t) > T_{\rm tri}$ and delineated by the 1/10-peak-width (FWTM) via linear interpolation to achieve sub-cadence temporal precision. Both the code and the data are available at Science Data Bank: \dataset[doi: 10.57760/sciencedb.36845]{\doi{10.57760/sciencedb.36845}}}\label{fig}   
\end{figure*}

\begin{figure*}
    \addtocounter{figure}{-1}
    \centering
    \newcommand{\figwidth}{0.46\linewidth}   
    \newcommand{\figsep}{\hspace{0.02\linewidth}} 
    
    \includegraphics[width = \figwidth, trim = 0 0 0 0, clip]{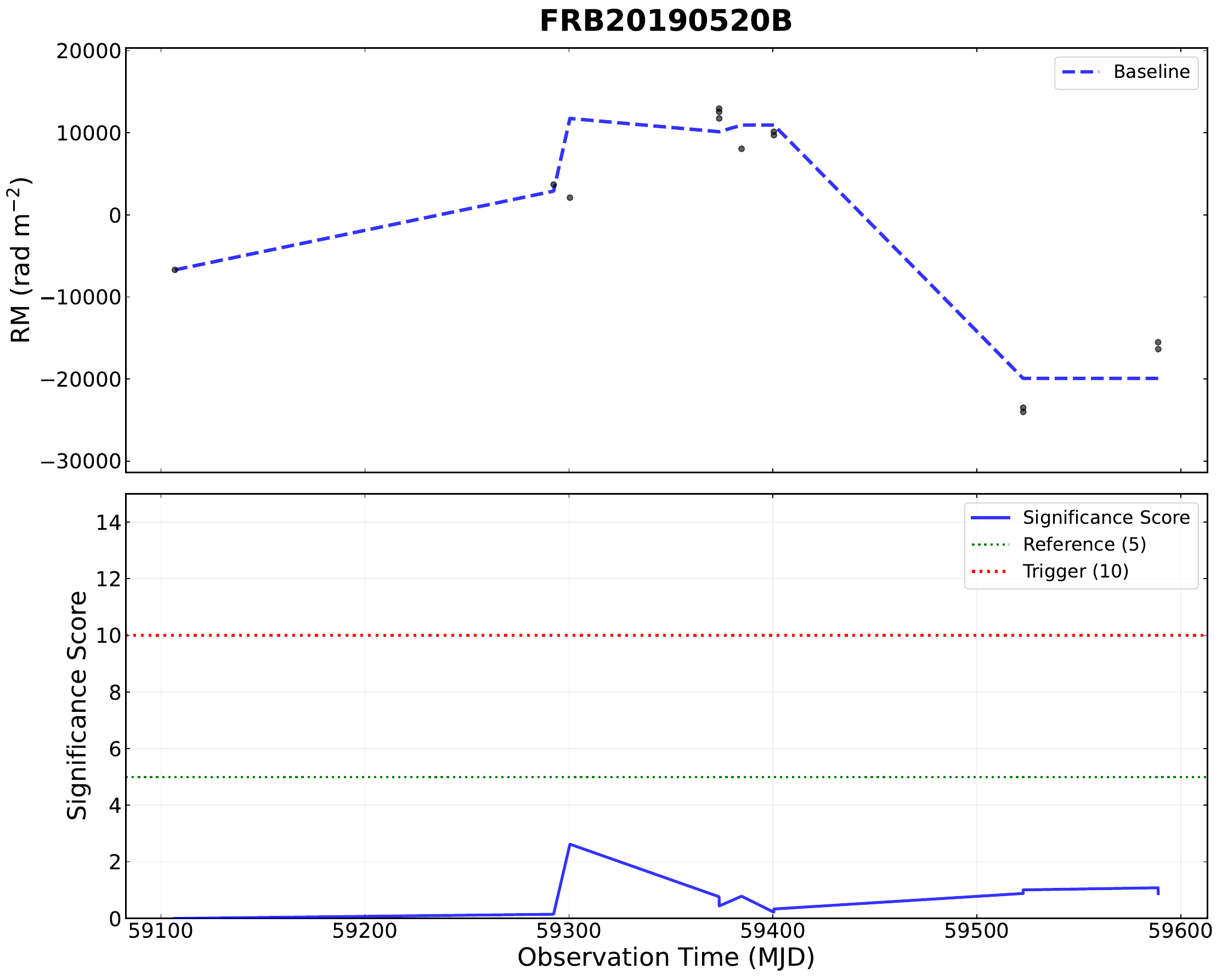} \figsep
    \includegraphics[width = \figwidth, trim = 0 0 0 0, clip]{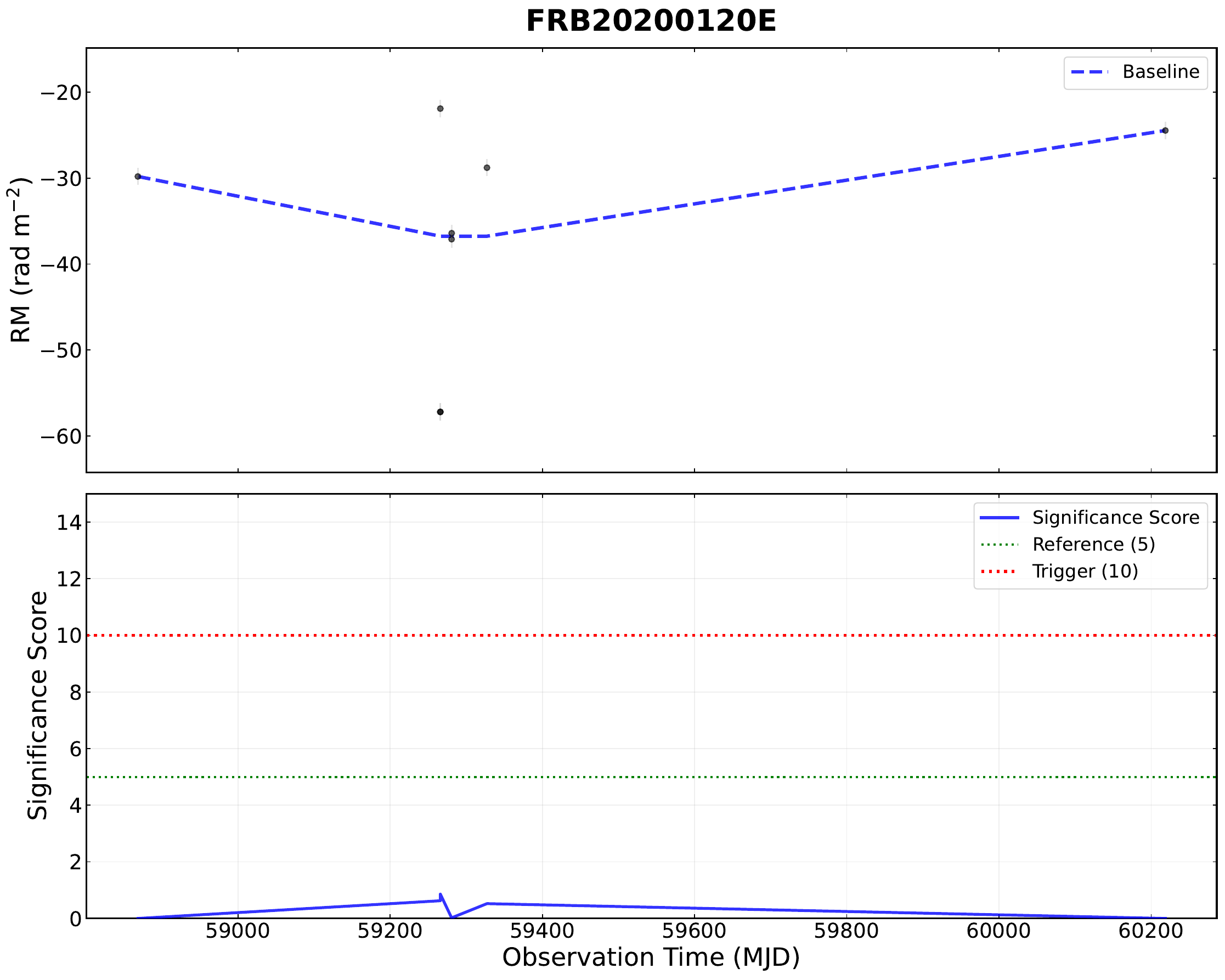} \\ \vspace{0.3cm}
    
    \includegraphics[width = \figwidth, trim = 0 0 0 0, clip]{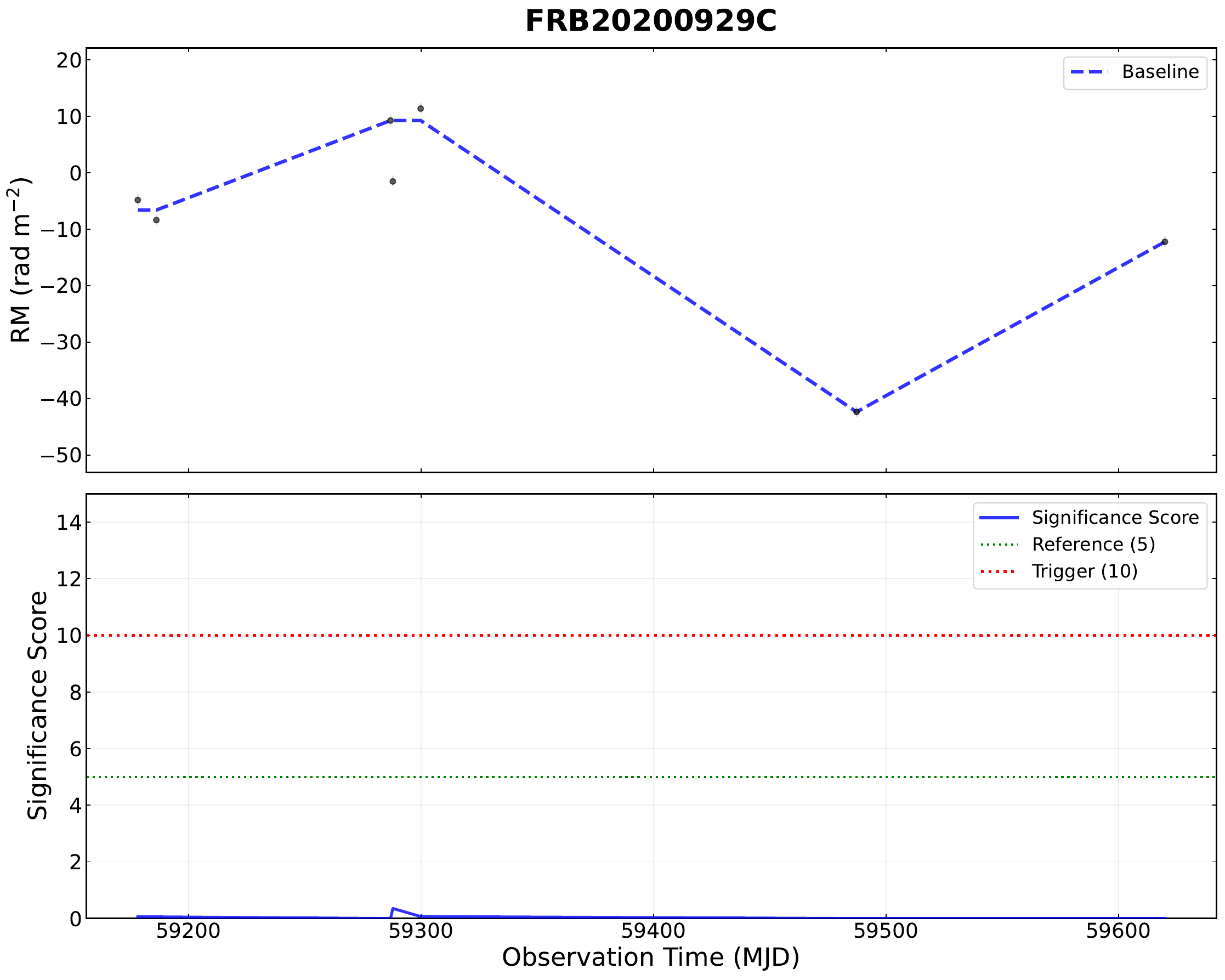} \figsep
    \includegraphics[width = \figwidth, trim = 0 0 0 0, clip]{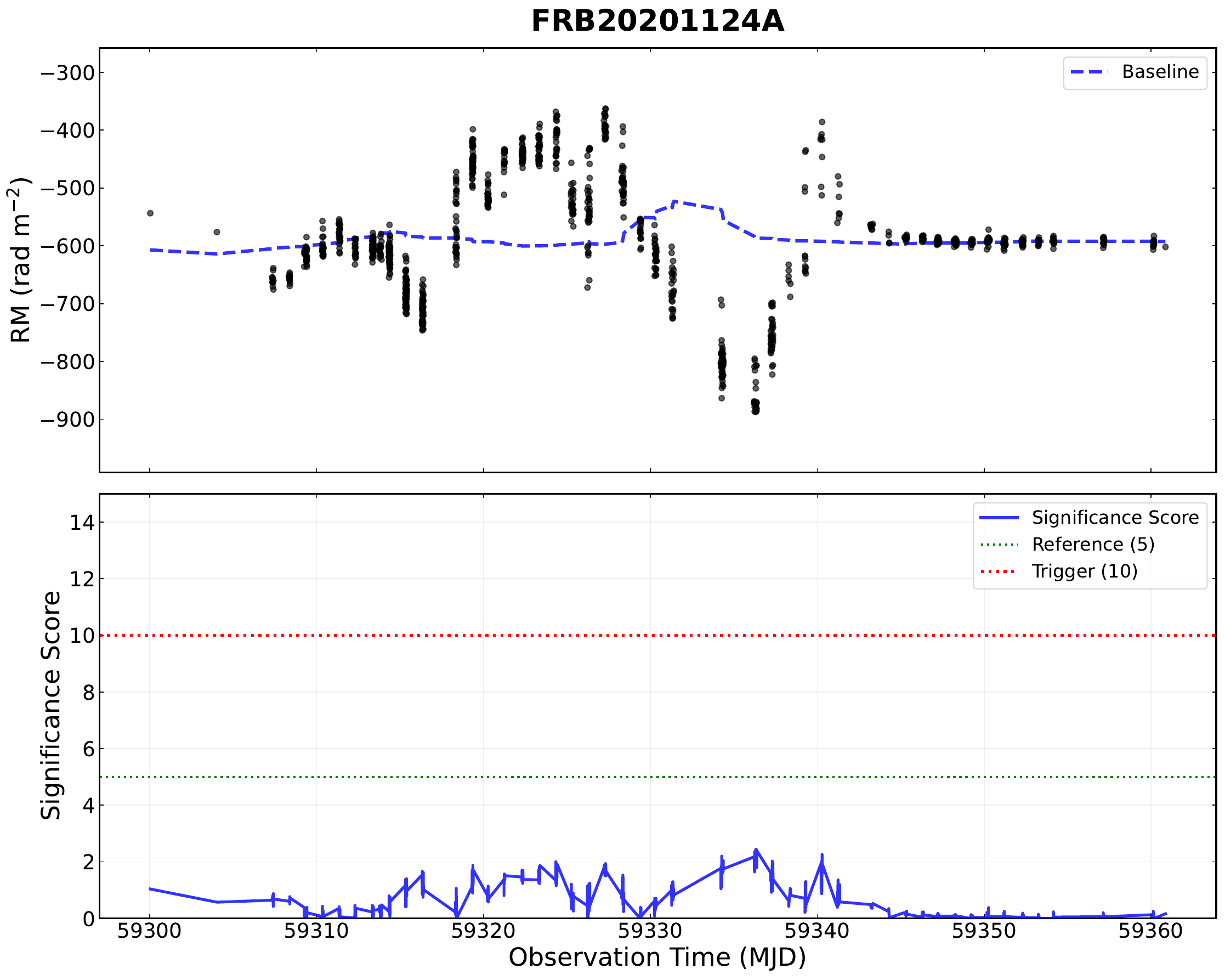} \\ \vspace{0.3cm}
    
    \includegraphics[width = \figwidth, trim = 0 0 0 0, clip]{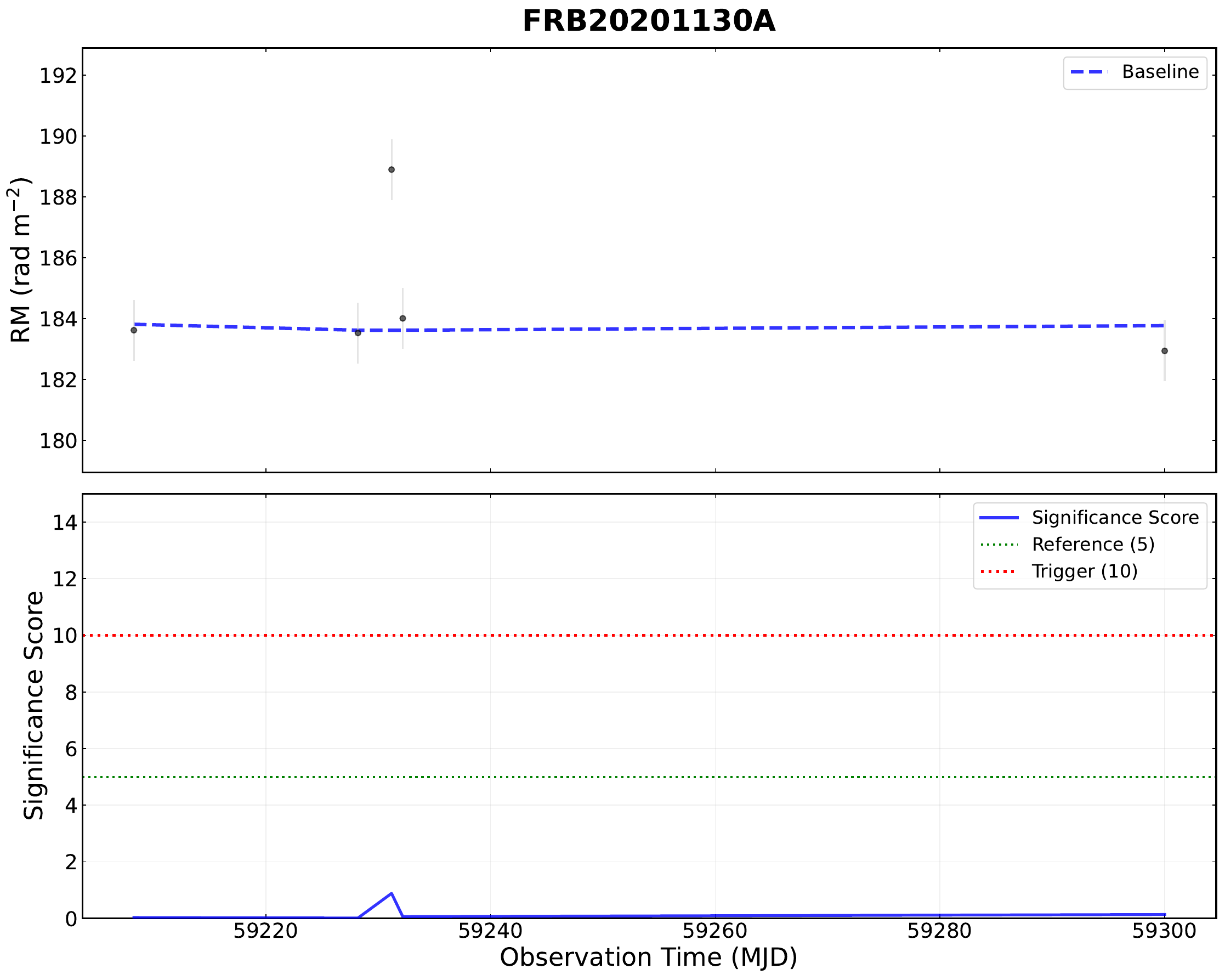} \figsep
    \includegraphics[width = \figwidth, trim = 0 0 0 0, clip]{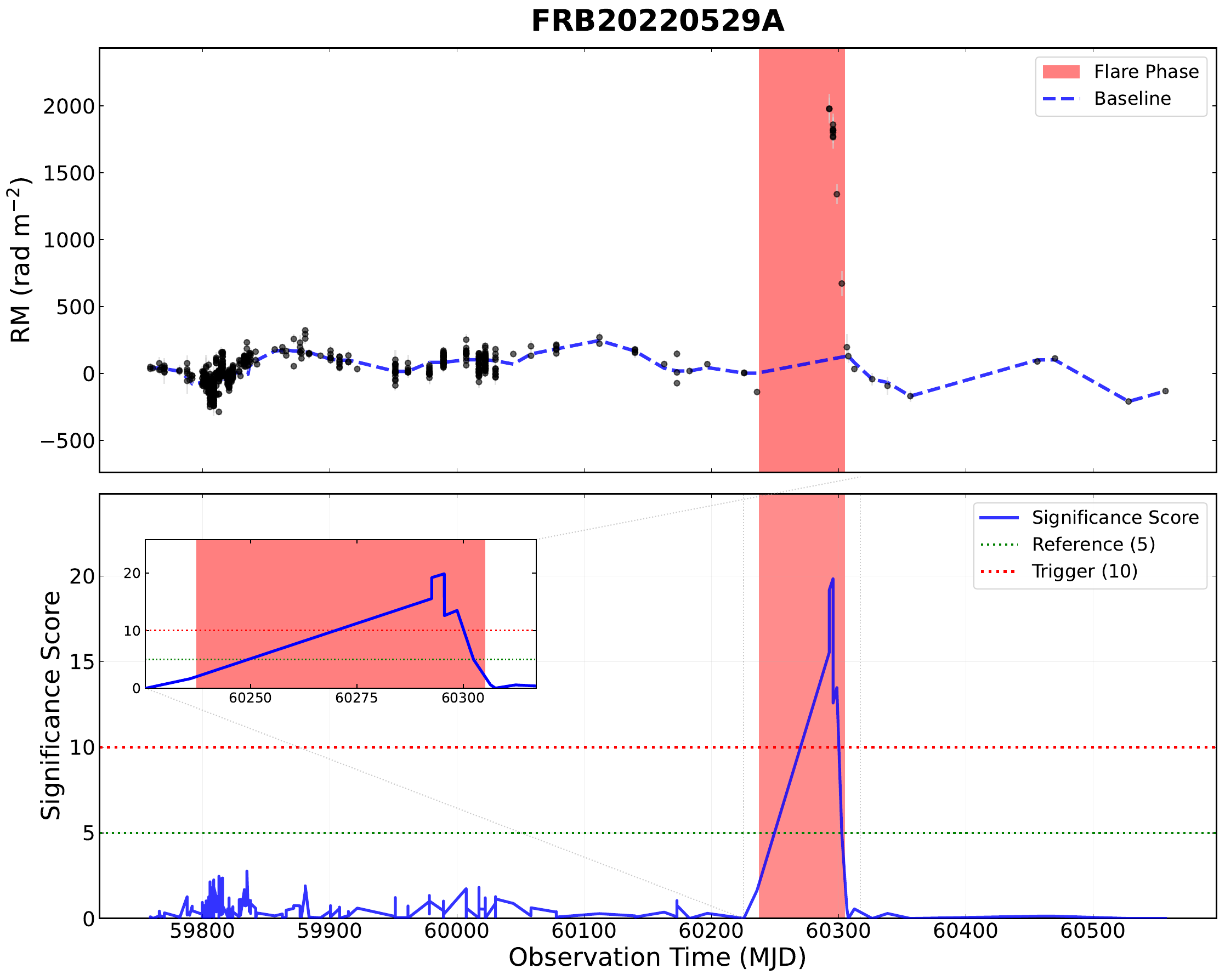}
    
    \caption{--- \textit{Continued.}}
\end{figure*}

\begin{figure*}
    \addtocounter{figure}{-1}
    \centering
    \newcommand{\figwidth}{0.46\linewidth}   
    \newcommand{\figsep}{\hspace{0.02\linewidth}} 
    
    \includegraphics[width = \figwidth, trim = 0 0 0 0, clip]{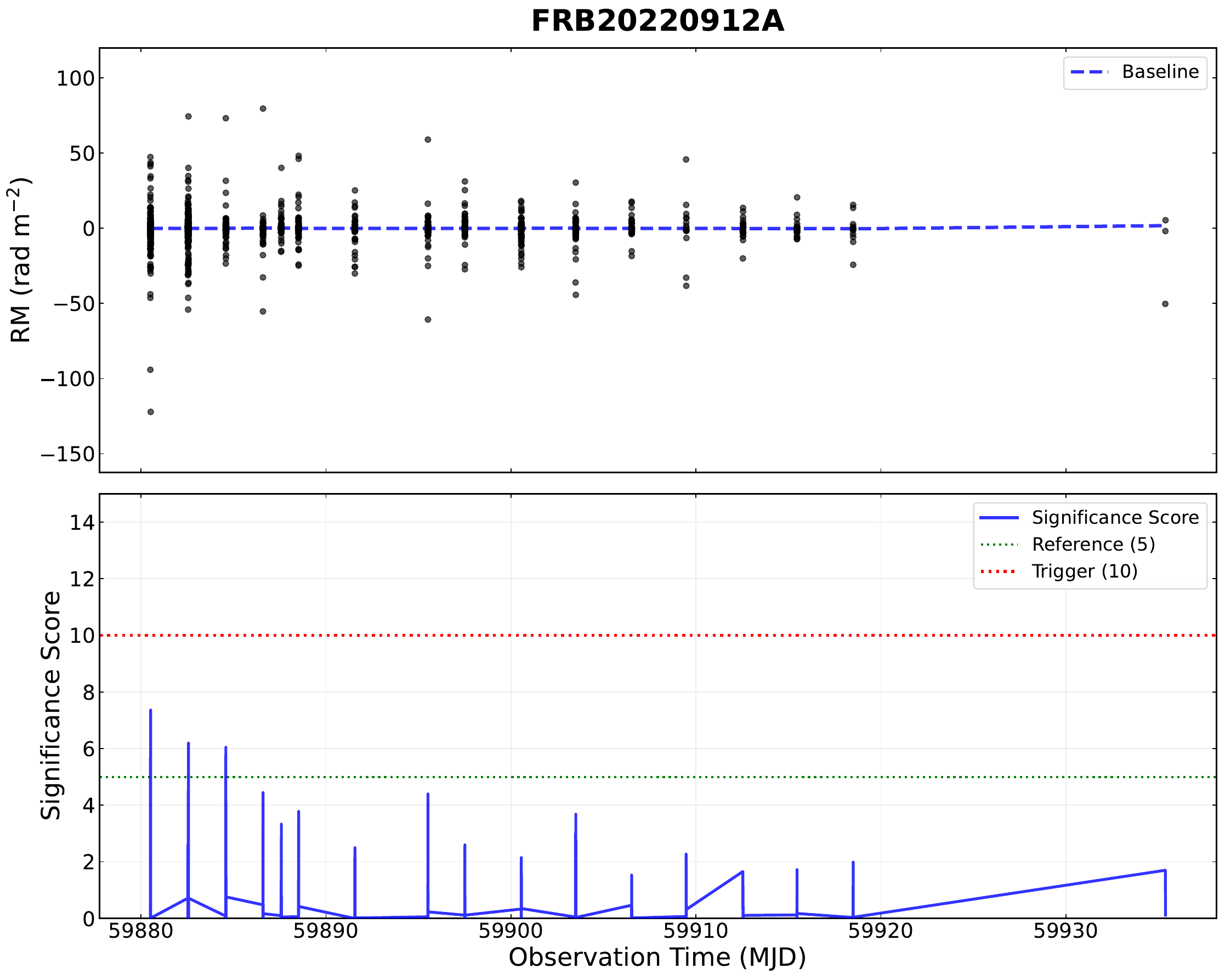} \figsep
    \includegraphics[width = \figwidth, trim = 0 0 0 0, clip]{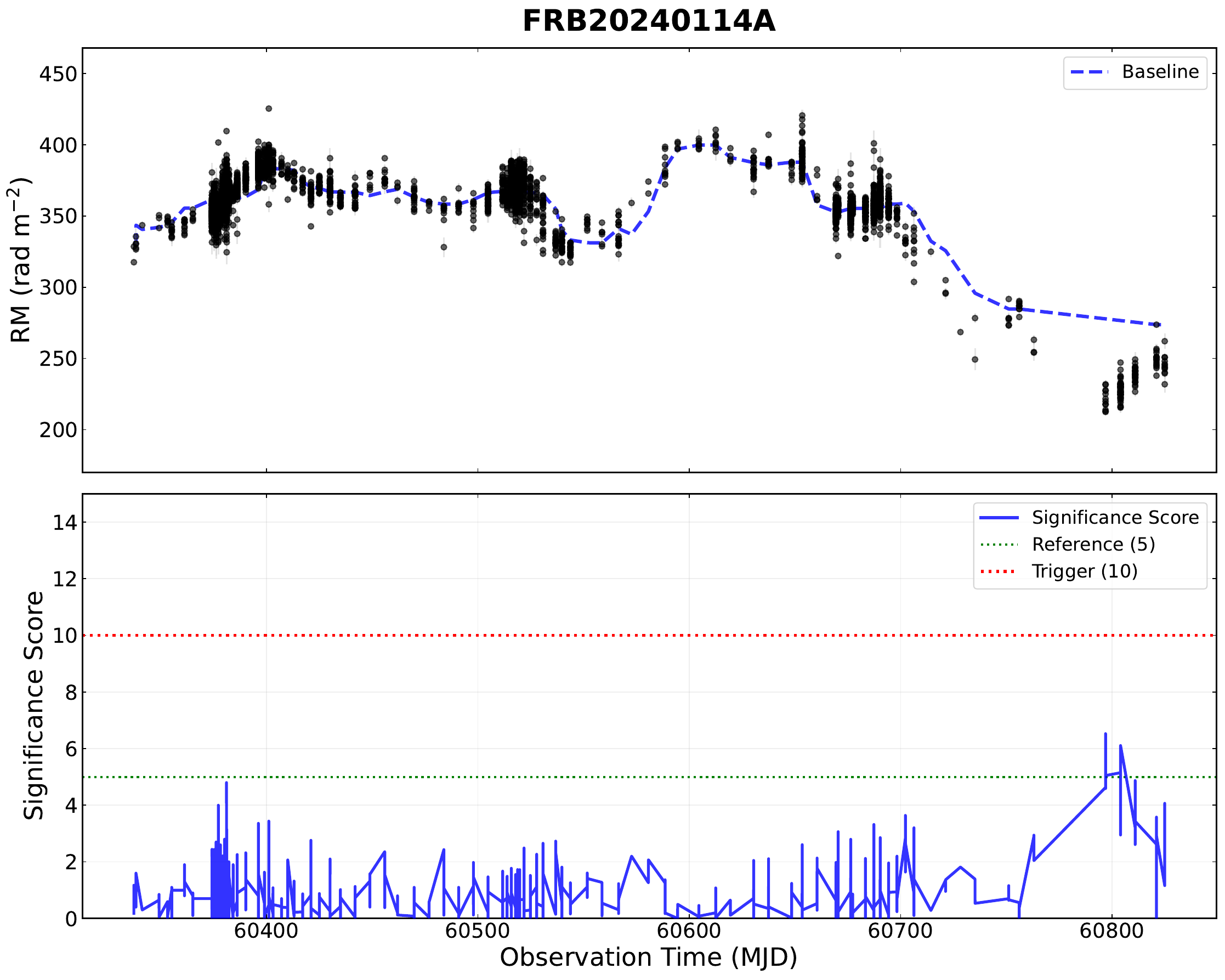} \\ \vspace{0.3cm}
    
    \includegraphics[width = \figwidth, trim = 0 0 0 0, clip]{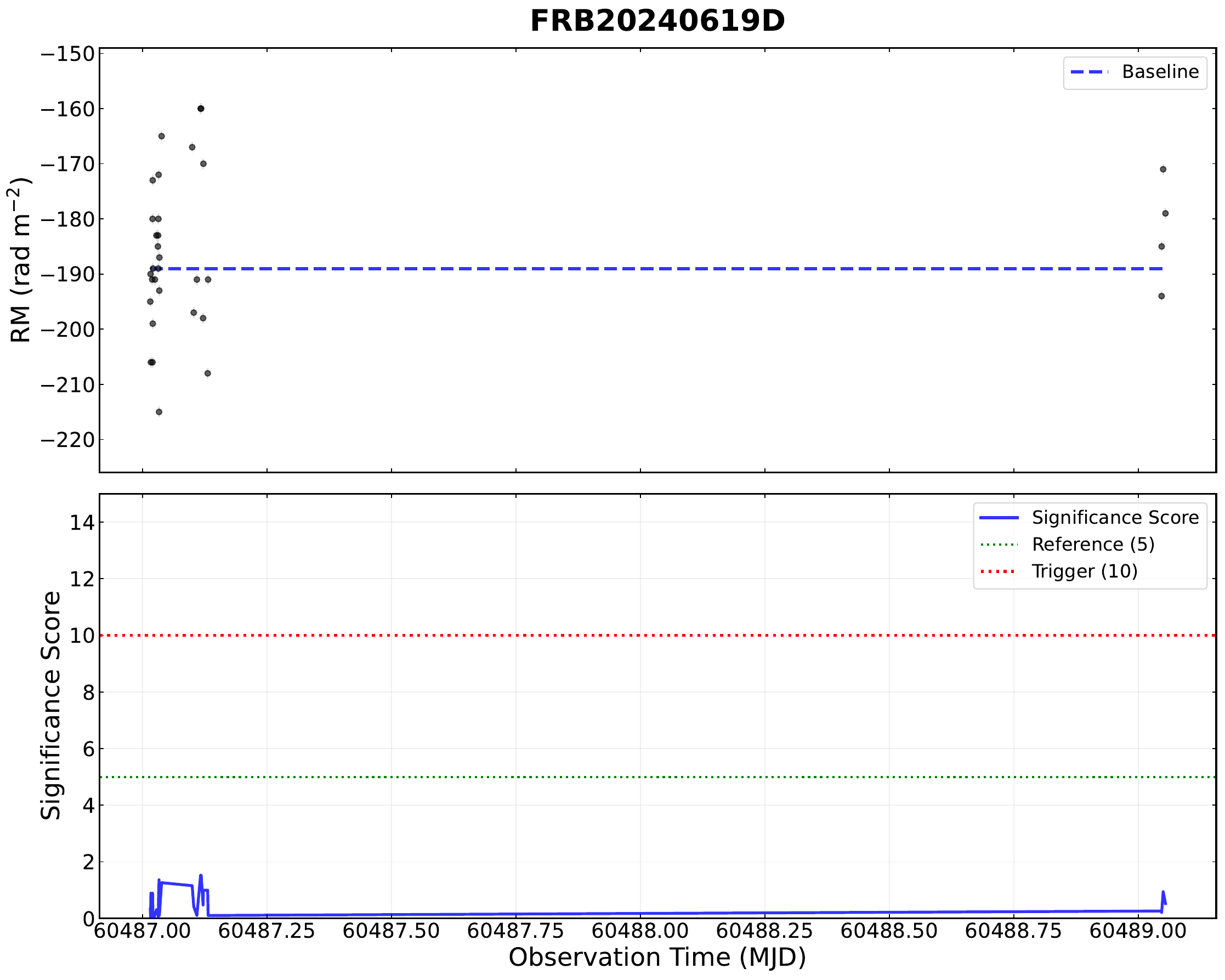} \figsep
    \makebox[\figwidth]{} 
    
    \caption{--- \textit{Continued.}}
\end{figure*}

We applied the automated detection pipeline to a sample of 15 repeating FRBs that have historically exhibited significant RM variations. The FRB sample is collected from the blinkverse database\footnote{\url{https://blinkverse.zero2x.org/}} \citep{Xu23}, while the RM evolution data for FRB 20220529A and FRB 20240114A are taken from \citet{Li26} and \citet{Wang26}, respectively. To maintain data consistency across diverse reporting standards, we implement a hierarchical selection criterion: the synthesized rotation measure (${\rm RM}_{\rm syn}$) is prioritized for analysis; in instances where ${\rm RM}_{\rm syn}$ is unavailable, values derived from QU-fitting (${\rm RM}_{\rm QUfit}$) are utilized as the fallback. By employing standardized configuration parameters (trigger threshold $T_{\rm tri} = 10$, width ratio $\eta = 0.1$, and scaling multiplier $k_{w} = 25$), we systematically evaluated the stability of their RM environments. 

The integrated results are presented in Figure \ref{fig}. For each source, the upper sub-panel displays the physical RM evolution, where black circles with error bars represent the observed values and the blue dashed line indicates the adaptive baseline $B(t)$. For FRB 20220529A, the identified flare phase is highlighted by the red shaded region. The lower sub-panel quantifies the significance through the Significance Score (blue solid line). To enhance the clarity of detection details, an inset zoom-in plot is provided for sources triggering a flare, focusing on the region surrounding the peak significance.

Our analysis demonstrates that while stochastic RM variability is a widespread trait, FRB 20220529A is the only source to satisfy the rigorous algorithmic criteria with  $S>T_{\rm tri} = 10$ for a discrete ``RM flare''. Notably, while FRB 20121102A, FRB 20220912A, and FRB 20240114A exhibit peak scores that slightly exceed the reference threshold $T_{\rm ref} = 5$ and visually resembles a flare-like excursion, they fail to reach the required trigger level of $T_{\rm tri} = 10$, indicating a lower confidence under our standardized parameters. Other active sources exhibit substantial absolute RM changes, but their scores remain well below the threshold or are interpreted as high-level intrinsic fluctuations due to the lack of a localized, peak-scaled structure. This automated population-wide scan ensures that the flare in FRB 20220529A is identified as a robust unique event, enabling systematic studies of local plasma dynamics. 

\section{Discussion and Conclusion}\label{conclusions}

The primary scientific value of this generalized framework lies in its ability to distinguish discrete, high-amplitude ``RM flares'' from widespread stochastic RM volatility. Our population-wide census confirms that genuine transient flares are rare, as the algorithm successfully isolates such unique events from widespread chaotic fluctuations and sub-threshold excesses. This observed rarity suggests that the physical conditions necessary to produce localized RM spikes, such as the passage of dense magneto-ionic clumps or stellar coronal mass ejections, are either intrinsically infrequent or require specific, seldom-realized line-of-sight geometries. By shifting the field from opportunistic, manual discovery to a systematic and reproducible census, our methodology establishes a rigorous empirical foundation for identifying transient physical interactions within the immediate magneto-ionic environments of FRB progenitors.

Furthermore, the parameterized architecture of the pipeline ($T_{\rm tri}$, $\eta$, $k_{w}$) ensures flexibility for diverse scientific objectives. While we employed a conservative trigger threshold ($T_{\rm tri} = 10$) and a sensitive width ratio ($\eta = 0.1$) to prioritize high-confidence detections, these parameters can be tuned to conduct broader population surveys of magneto-ionic stability. As polarization data from facilities like FAST, CHIME, and the SKA continue to accumulate, this standardized analysis tool will be essential for managing vast datasets and correlating RM flares with other burst properties, such as activity cycles, spectral changes, or persistent radio source associations. 

\section*{Acknowledgments}
We thank the anonymous referee for providing helpful comments and suggestions.
We also acknowledge Ye Li for stimulating discussions that motivated the pursuit of this research.
This work is supported by the National Natural Science Foundation of China (No. 12473047), the National Key Research and Development Program of China (No. 2024YFA1611603) and the Yunnan Key Laboratory of Survey Science (No. 202449CE340002). 

\emph{Code Availability}:
The Python implementation of our RM-flare detection algorithm, along with the simulated and observational datasets used in this work, has been deposited in a public repository with an assigned DOI: \url{https://doi.org/10.57760/sciencedb.36845}

\appendix

\section{Algorithm Pseudocode of Automated RM Flare Detection Pipeline}\label{pseudocode}

This appendix presents the step-by-step algorithmic pseudocode for the automated RM flare detection pipeline introduced in Section 2. It details the sequential logic, dynamic masking conditions, and parameter operations required to extract discrete flare signatures from complex background evolution.

\begin{figure}[H]
\hrule \vspace{2mm}
\noindent \textbf{Algorithm:} Automated RM Flare Detection Pipeline\\
\vspace{-2mm} \hrule \vspace{2mm}
\begin{algorithmic}[1]
\Require 
    Observation epochs $t$, Rotation Measures ${\rm RM}(t)$, Formal errors $\sigma_{\rm err}(t)$
\Require 
    Hyperparameters: $k_w$ (window multiplier), $(W_{\rm min},W_{\rm max})$ (window limits), $(N_{\rm glob},N_{\rm loc})$ (screening thresholds), $\eta$ (width ratio), $T_{\rm tri}$ (trigger threshold)

\Statex \textbf{Stage 1: Adaptive Windowing \& Global Screening}
\State $t_{\rm unique} \gets \text{Unique observing epochs from } t$ \Comment{Extract epochs for cadence calculation}
\State $\Delta t_{\rm med} \gets \text{Median}(|t_{\rm unique, i} - t_{\rm unique, i-1}|)$ \Comment{Median observational cadence}
\State $W \gets \text{clip}(k_w \times \Delta t_{\rm med},\ W_{\rm min},\ W_{\rm max})$ \Comment{Adaptive temporal window size}
\State $\text{Median}_{\rm glob} \gets \text{Median}({\rm RM}(t))$ \Comment{Global median of original RM}
\State $\text{MAD}_{\rm glob} \gets \text{Median}(|{\rm RM}(t) - \text{Median}_{\rm glob}|)$ \Comment{Global Median Absolute Deviation}
\State Calculate global bias $\Delta_{\rm RM}$ and local bias $\delta_{\rm RM}$ based on ${\rm RM}(t)$ \Comment{Data-driven bias constraints}
\State $M_{\rm extreme} \gets \text{Indices where } |{\rm RM}(t) - \text{Median}_{\rm glob}| > (N_{\rm glob} \times \text{MAD}_{\rm glob} + \Delta_{\rm RM})$ 
\State $M_{\rm quiescent} \gets \text{All valid indices} \setminus M_{\rm extreme}$ \Comment{Initial mask for quiescent epochs}

\Statex \textbf{Stage 2: Iterative Local Baseline Estimation on Full Cadence}
\Repeat
    \State $B_{\rm roll} \gets \text{RollingMedian}({\rm RM}[M_{\rm quiescent}], \text{window}=W)$ \Comment{Rolling baseline on quiescent subset}
    \State $B_{\rm temp}(t) \gets \text{Linearly interpolate } B_{\rm roll} \text{ back to all original epochs } t$ \Comment{Continuous baseline}
    \State $\sigma_{\rm roll} \gets \text{RollingMedian}(|{\rm RM}[M_{\rm quiescent}] - B_{\rm temp}(t[M_{\rm quiescent}])|, \text{window}=W)$ 
    \State $\sigma_{\rm temp}(t) \gets \text{Linearly interpolate } \sigma_{\rm roll} \text{ back to all original epochs } t$
    \State $\sigma_{\rm temp}(t) \gets \max(\sigma_{\rm temp}(t), \sigma_{\rm floor})$ \Comment{Apply dynamic instrumental noise floor}
    \State $M_{\rm candidate} \gets \text{Indices where } |{\rm RM}(t) - B_{\rm temp}(t)| < (N_{\rm loc} \times \sigma_{\rm temp}(t) + \delta_{\rm RM})$ 
    \State $M_{\rm new} \gets M_{\rm candidate} \setminus M_{\rm extreme}$ \Comment{Exclude extreme outliers before convergence check}
    \State \textbf{if} $M_{\rm new} == M_{\rm quiescent}$ \textbf{then} Converged $\gets$ True
    \State $M_{\rm quiescent} \gets M_{\rm new}$
\Until{Converged \textbf{or} Max Iterations reached}
\State $B(t) \gets B_{\rm temp}(t)$ \Comment{Final continuous baseline}
\State $\sigma_{\rm loc}(t) \gets \sigma_{\rm temp}(t)$ \Comment{Final continuous local noise}

\Statex \textbf{Stage 3: Multi-Component Noise Modeling \& Significance Scoring}
\State Calculate $\sigma_{\rm intra}(t)$ \Comment{Intra-day scatter derived from original multi-burst epochs}
\State $\mathcal{R}(t) \gets |{\rm RM}(t) - B(t)|$ \Comment{Absolute residuals at original high resolution}
\State $\sigma_{\rm glob} \gets \text{Median}(|\mathcal{R}(t) - \text{Median}(\mathcal{R}(t))|)$ \Comment{Intrinsic global volatility floor}
\State $\sigma_{\rm tot}(t) \gets \sqrt{\sigma_{\rm glob}^2 + \sigma_{\rm loc}(t)^2 + \sigma_{\rm intra}(t)^2 + \sigma_{\rm err}(t)^2}$ \Comment{Total effective noise}
\State $S(t) \gets \mathcal{R}(t) \,/\, \max(\sigma_{\rm safe}, \sigma_{\rm tot}(t))$ \Comment{Empirical pseudo-SNR, bounded by safety floor}

\Statex \textbf{Stage 4: Flare Trigger \& Physical Boundary Interpolation}
\State $\text{Candidate Segments} \gets \text{Contiguous groups where } S(t) > 0.5$
\For{\textbf{each} segment \textbf{in} Candidate Segments}
    \State $S_{\rm peak} \gets \max(S(t) \text{ in segment})$ \Comment{Peak score of the current segment}
    \If{$S_{\rm peak} \ge T_{\rm tri}$} \Comment{Evaluate against rigorous trigger threshold}
        \State $t_{\rm peak} \gets \text{Epoch corresponding to } S_{\rm peak}$
        \State $t_{\rm start} \gets \text{Linear interpolation of } t \text{ backward from } t_{\rm peak} \text{ where } S(t) = \eta S_{\rm peak}$ \Comment{Left FWTM boundary}
        \State $t_{\rm end} \gets \text{Linear interpolation of } t \text{ forward from } t_{\rm peak} \text{ where } S(t) = \eta S_{\rm peak}$ \Comment{Right FWTM boundary}
        \State \textbf{Save} Flare Interval $[t_{\rm start}, t_{\rm end}]$
    \EndIf
\EndFor
\State \textbf{Return} $B(t)$, $S(t)$, and Flare Intervals
\end{algorithmic}
\vspace{2mm} \hrule
\vspace{2mm} 
\end{figure}

\section{Algorithm Verification via Controlled Tests}
\label{sec:appendix_mock} 

\begin{figure*}[htbp]
    \centering
    \includegraphics[width = 0.9\linewidth, trim = 0 0 0 0, clip]{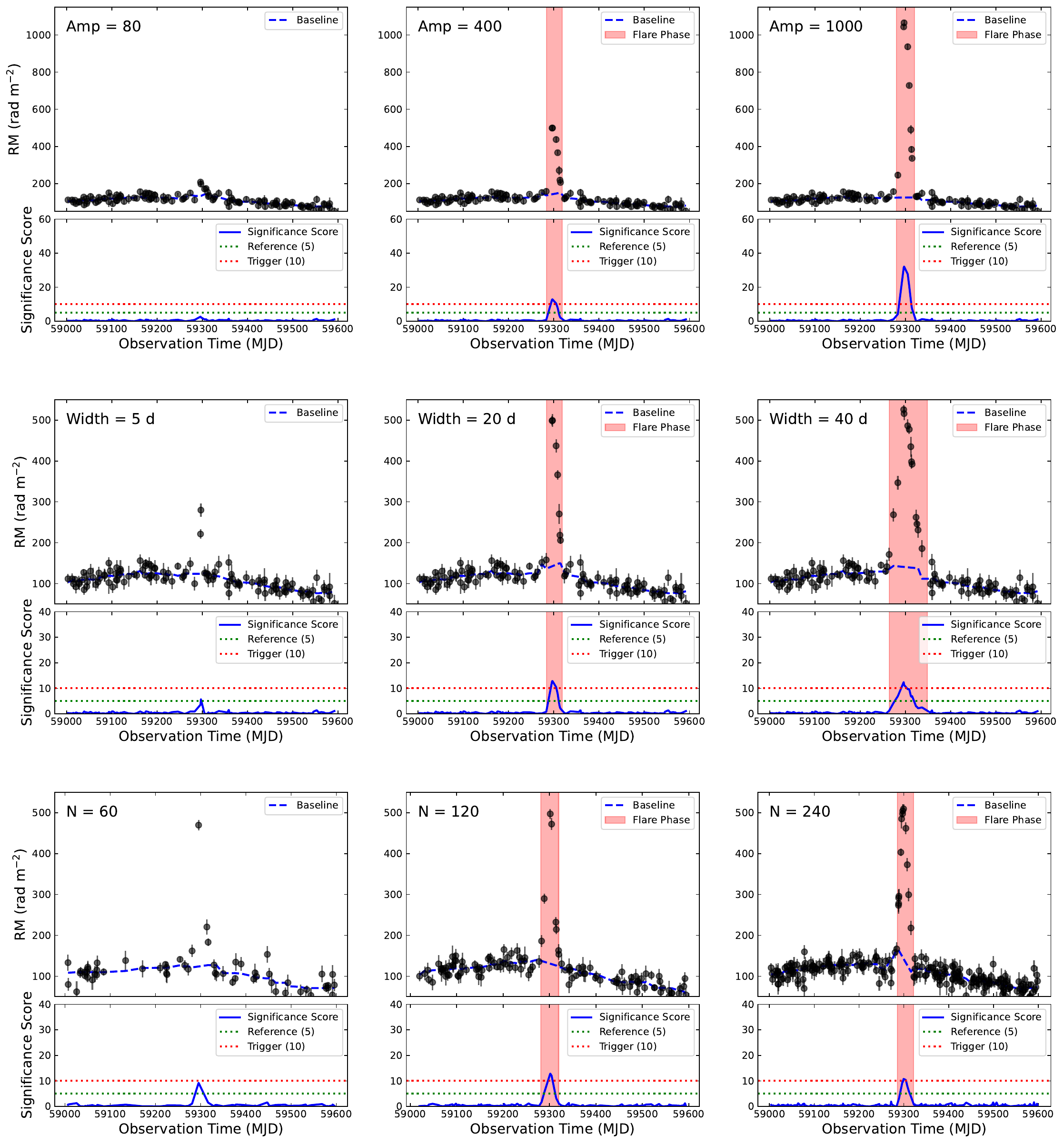}
    \caption{Controlled tests evaluating the algorithm's performance across varying flare properties and observational cadences. The figure is divided into three sets of tests (from top to bottom): Varying Amplitude (fixed FWHM of 20 days), Varying Duration (fixed Amplitude of 400 ${\rm rad~m^{-2}}$), and Varying Sampling Density (fixed Amplitude of 400 ${\rm rad~m^{-2}}$ and FWHM of 20 days). In each set, the upper sub-panels show the simulated RM data (black markers) and the algorithmically derived adaptive baseline (blue dashed line). The red shaded regions represent the identified ``Flare Phase''. The lower sub-panels display the computed Significance Score (blue solid line), plotted against the reference threshold of $T_{\rm ref}=5$ (green dotted line) and the rigorous trigger threshold of $T_{\rm tri}=10$ (red dotted line).}
    \label{fig:injection_tests}
\end{figure*}

To quantitatively validate the proposed algorithm and evaluate its detection sensitivity and robustness, we conduct a series of controlled tests using simulated RM time series. The synthetic quiescent background is modeled to mimic the complex secular evolution and stochastic noise typical of repeating FRBs. 
Specifically, the background RM evolution is defined as ${\rm RM}_{\rm base}(t) = {\rm RM}_0 + A_{\rm sec} \sin(2\pi(t - t_{\rm base})/P_{\rm sec}) + \mathcal{N}(0, \sigma_{\rm err})$, where the baseline constant ${\rm RM}_0 = 100~{\rm rad~m^{-2}}$, the secular amplitude $A_{\rm sec} = 30~{\rm rad~m^{-2}}$, and the long-term period $P_{\rm sec} = 800$ days. The measurement errors, $\sigma_{\rm err}$, are drawn from the absolute values of a normal distribution, $|\mathcal{N}(15.0, 5.0)|$. Upon this background, we inject a discrete Gaussian flare at a central epoch $t_0$, defined by ${\rm RM}_{\rm obs}(t) = {\rm RM}_{\rm base}(t) + A_{\rm flare} \exp(-(t - t_0)^2/2\sigma_{\rm flare}^2)$, where $A_{\rm flare}$ is the peak amplitude and $\sigma_{\rm flare}$ controls the duration (converted from the Full Width at Half Maximum via $\sigma_{\rm flare} = {\rm FWHM}/2.355$). We systematically evaluate the algorithm's detection performance and its operational boundaries across a range of flare amplitudes, durations, and observational cadences, as illustrated in Figure \ref{fig:injection_tests}. 

\begin{itemize}
    \item \textbf{Varying Amplitude:} To evaluate the detection sensitivity, we first test a scenario with varying flare amplitudes (fixed ${\rm FWHM} = 20$ days) on a normally sampled dataset ($N = 120$). As shown in the top panels of Figure \ref{fig:injection_tests}, we inject flares with amplitudes of 80, 400, and 1000 ${\rm rad~m^{-2}}$. For the weak perturbation ($A_{\rm flare}=80~{\rm rad~m^{-2}}$, top-left), the peak Significance Score reaches approximately 4, remaining strictly below the trigger threshold of $T_{\rm tri} = 10$ (red dotted line). Consequently, no flare phase is triggered, demonstrating the algorithm's conservative stance against low-amplitude fluctuations. Conversely, for the clear (400) and massive (1000) flares, the scores confidently exceed the threshold. The dynamic baseline (blue dashed line) is unperturbed by the extreme outliers, ensuring the flare's significance is evaluated without signal self-subtraction.

    \item \textbf{Varying Duration:} To assess the algorithmic response to different event morphologies, we conduct a second suite of tests by fixing the amplitude at $400~{\rm rad~m^{-2}}$ and varying the FWHM to simulate a sharp spike (5 days), a standard flare (20 days), and a broad bump (40 days). In the middle-left panel, the sharp 5-day spike fails to trigger the detection. Given the normal sampling cadence ($N=120$ over 600 days), such a brief transient lacks sufficient contiguous data points to build up a robust score above $T_{\rm tri} = 10$. For the 20-day and 40-day flares, the algorithm successfully triggers and dynamically bounds the ``Flare Phase'' (red shaded regions). Notably, even for the 40-day broad bump, the algorithm bounds the event without absorbing the long-duration transient into the secular baseline.

    \item \textbf{Varying Sampling Density:} Finally, we test the algorithm's dependency on observational cadence by injecting an identical flare ($A_{\rm flare}=400~{\rm rad~m^{-2}}$, ${\rm FWHM} = 20$ days) into sparse ($N = 60$), normal ($N = 120$), and dense ($N = 240$) temporal grids. As depicted in the bottom panels of Figure \ref{fig:injection_tests}, the sparse sampling (bottom-left) degrades the peak score to around 6, failing to meet the rigorous $T_{\rm tri} = 10$ requirement. This visually confirms that insufficient temporal sampling cannot provide the required confidence for a discrete flare detection under our standardized criteria. In contrast, for the normal and dense datasets, the flare is identified, and the adaptive baseline correctly scales with the data density without overfitting the intra-day noise. 
\end{itemize}

These controlled tests explicitly map the operational boundaries of our detection framework. The algorithm successfully identifies and bounds distinct, sufficiently sampled RM flares, while remaining highly conservative against sparse, weak, or excessively narrow perturbations.

\section{Parameter Sensitivity Analysis}
\label{sec:appendix_sensitivity}

\begin{figure*}
    \centering
    \includegraphics[width = 0.9\linewidth, trim = 0 0 0 0, clip]{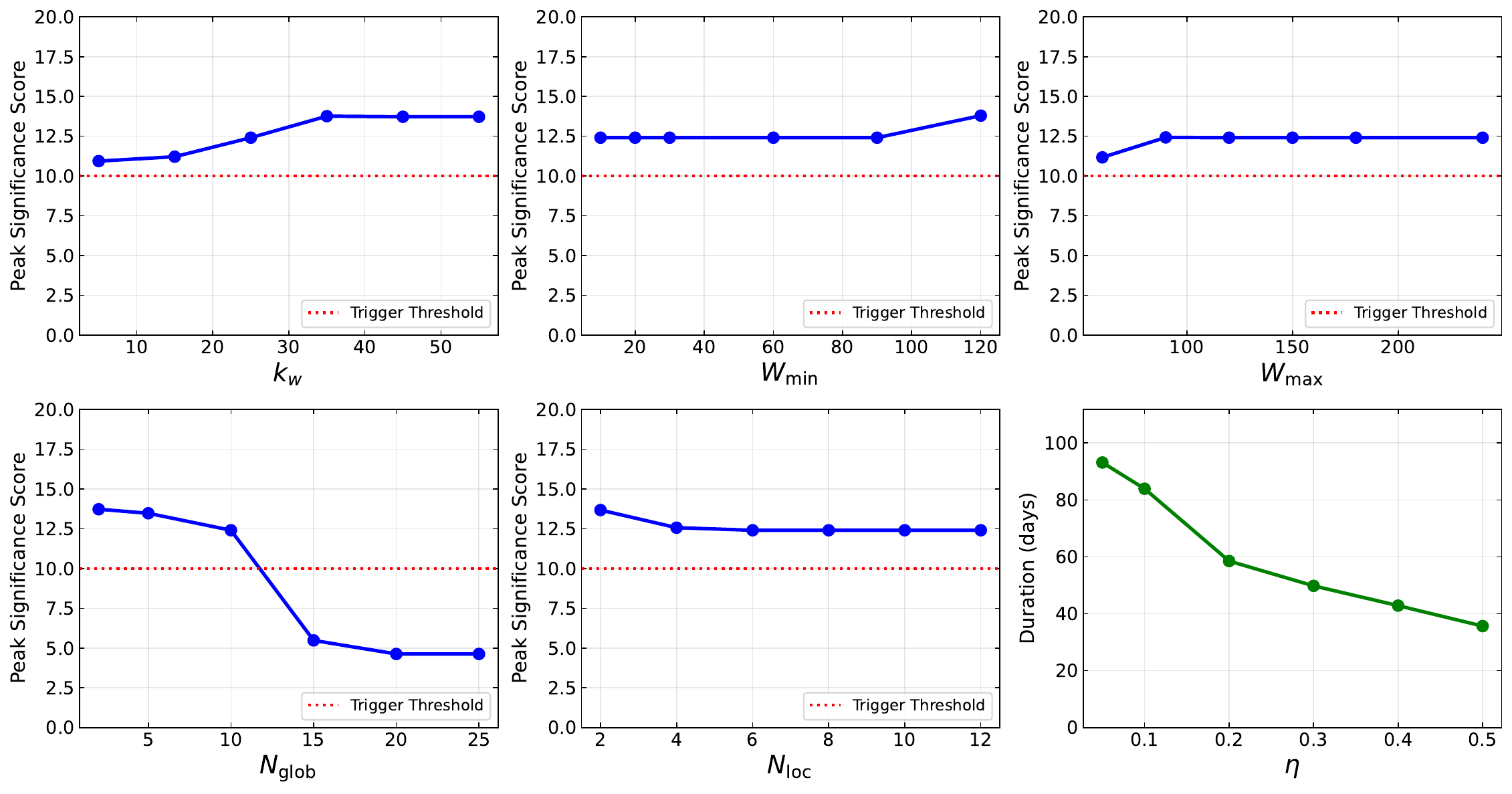}
    \caption{Parameter sensitivity analysis of the automated RM flare detection framework. We tested the impact of the scaling multiplier ($k_w$), adaptive window boundaries ($W_{\rm min}$, $W_{\rm max}$), global/local screening thresholds ($N_{\rm glob}$, $N_{\rm loc}$), and the width ratio ($\eta$). The first five sub-panels display the Peak Significance Score (blue line) as a function of the parameter value, overlaid with the rigorous trigger threshold $T_{\rm tri}=10$ (red dotted line). The bottom-right sub-panel shows the extracted flare duration (green line) as a function of $\eta$.}
    \label{fig:sensitivity}
\end{figure*} 

To systematically justify the choices of the algorithmic parameters and evaluate the stability of our detection pipeline, we performed a parameter sensitivity analysis. As pointed out, the algorithm depends on several key hyperparameters: the window scaling multiplier ($k_w$), the adaptive window boundaries ($W_{\rm min}$, $W_{\rm max}$), the global and local screening thresholds ($N_{\rm glob}$, $N_{\rm loc}$), and the width ratio ($\eta$). We aim to demonstrate that our standardized configuration does not strictly rely on finely tuned parameter choices, though certain parameters exhibit non-negligible sensitivity. Nevertheless, it remains within a reasonably broad and relatively stable operational regime.

To pose a rigorous challenge to the algorithm, we utilized a mock background injected with a broad flare (normal sampling $N=120$, injected amplitude $= 400~{\rm rad~m^{-2}}$, and FWHM $= 40$ days). We systematically varied one parameter at a time across a wide dynamic range while keeping the others fixed at their default standardized values ($k_w=25$, $W_{\rm min}=30$, $W_{\rm max}=150$, $N_{\rm glob}=10$, $N_{\rm loc}=5$, $\eta=0.1$). To quantify the algorithmic performance, we extracted the Peak Significance Score for the first five parameters to assess trigger stability, and the recovered Flare Duration for $\eta$ to assess the boundary extraction logic. The results are presented in Figure \ref{fig:sensitivity}.

The sensitivity analysis reveals several critical algorithmic behaviors:

\begin{itemize}
    \item \textbf{Adaptive Windowing Parameters ($k_w$, $W_{\rm min}$, $W_{\rm max}$):} The Peak Significance Score exhibits absolute stability across a vast range of lower ($W_{\rm min}$) and upper ($W_{\rm max}$) boundaries, remaining consistently well above the rigorous trigger threshold ($T_{\rm tri}=10$). For the scaling multiplier $k_w$, the score experiences a slight reduction at small values ($k_w \le 30$), which is physically expected: an overly narrow rolling window begins to over-fit the broad flare profile. Nevertheless, the score remains above the detection threshold across the entire tested domain. 
    
    \item \textbf{Iterative Screening Thresholds ($N_{\rm glob}$, $N_{\rm loc}$):} These parameters govern the masking of extreme outliers to prevent them from polluting the background noise estimation. As shown in the bottom-middle panel of Figure \ref{fig:sensitivity}, the algorithm is highly robust against local filtering variations, maintaining a stable score across all tested $N_{\rm loc}$ values. For the global screening ($N_{\rm glob}$), the algorithm acts stably around our default parameter ($N_{\rm glob} \lesssim 10$). However, if this threshold is set too loose (e.g., $N_{\rm glob} \ge 12$), the 40-day broad flare is no longer fully masked during the initial stage. This forces the global noise floor ($\sigma_{\rm glob}$) to artificially inflate, thereby dragging the peak score below the trigger threshold. Our default value ($N_{\rm glob} = 10$) can isolate the quiescent background from such transient excursions.
    
    \item \textbf{Width Ratio ($\eta$):} Unlike the other parameters that affect the Peak Significance Score, $\eta$ explicitly defines the physical boundaries of the ``Flare Phase''. As $\eta$ increases from 0.05 to 0.5, the measured duration monotonically decreases, tracing the mathematical profile of the injected Gaussian flare. For instance, at $\eta=0.5$, the algorithm correctly recovers a duration of $\sim 40$ days, corresponding to the injected FWHM. Under our standardized definition of $\eta=0.1$ (FWTM), the theoretical duration for a Gaussian pulse is expected to be $\simeq 1.823 \times {\rm FWHM} \sim 70$ days. The algorithm extracts a duration of $\sim 84$ days. This slight extension is a natural consequence of incorporating the background noise scatter at the extremely low-intensity tails, demonstrating the algorithm's capability to safely encompass the RM flare phase.
\end{itemize}

In conclusion, the parameter choices in our generalized framework are reasonably well motivated. The analysis suggests that alternative choices within a plausible parameter space lead to broadly consistent flare detections, although some variation may arise under more extreme burst morphologies. The default configuration lies within a relatively stable and physically motivated regime, supporting a generally uniform and reproducible census across the repeating FRB population.


\bibliographystyle{aasjournal}

\end{document}